\documentclass[11pt,authoryear]{elsarticle}

\usepackage{amsmath,amssymb,amsthm,amsfonts}
\usepackage{mathabx}
\usepackage{mathrsfs}
\usepackage{bm}
\usepackage{graphicx}
\usepackage{float}
\usepackage{subfig}
\usepackage{color,xcolor}
\usepackage{tabularx}
\usepackage{epsfig,psfrag}
\usepackage{tikz}
\usepackage{pgflibraryshapes}
\usepackage{graphicx,psfrag}
\setcitestyle{authoryear,round}
\usepackage{hyperref}

\usepackage{changes}
\makeatletter
\@namedef{Changes@AuthorColor}{blue}
\colorlet{Changes@Color}{blue}
\makeatother

\makeatletter
\hypersetup{colorlinks=true}
\AtBeginDocument{\@ifpackageloaded{hyperref}
	{\def\@linkcolor{magenta}
		\def\@anchorcolor{black}
		\def\@citecolor{teal}
		\def\@filecolor{cyan}
		\def\@urlcolor{magenta}
		\def\@menucolor{red}
		\def\@pagecolor{cyan}
		\begingroup
		\@makeother\`%
		\@makeother\=%
		\edef\x{%
			\edef\noexpand\x{%
				\endgroup
				\noexpand\toks@{%
					\catcode 96=\noexpand\the\catcode`\noexpand\`\relax
					\catcode 61=\noexpand\the\catcode`\noexpand\=\relax
				}%
			}%
			\noexpand\x
		}%
		\x
		\@makeother\`
		\@makeother\=
	}{}}
\makeatother

\makeatletter \oddsidemargin 0.0cm \evensidemargin 0.0cm
\newcommand{\be}{\begin{equation}}
\newcommand{\en}{\end{equation}}
\newcommand{\la}{\label}

\newcommand{\muu}{{a}}
\newcommand{\muub}{{b}}
\newcommand{\paa}{\partial}
\def\rr#1{(\ref{#1})}
\def\bm#1{\mbox{\boldmath{$#1$}}}

\numberwithin{equation}{section}
\theoremstyle{plain}

\newtheorem{theorem*}{Theorem}

\theoremstyle{definition}

\DeclareMathOperator{\sech}{sech}

\journal{Journal of the Mechanics and Physics of Solids}

\begin{document}

\begin{frontmatter}

\title{{\bf A one-dimensional model for axisymmetric deformations of an inflated hyperelastic tube of finite wall thickness}}

\author[mymainaddress]{Xiang Yu\corref{mycorrespondingauthor}}
\cortext[mycorrespondingauthor]{Corresponding author}
\ead{yuxiang@dgut.edu.cn}

\author[mysecondaryaddress]{Yibin Fu}

\address[mymainaddress]{School of Computer Science and Technology, Dongguan University of Technology, Dongguan, China}
\address[mysecondaryaddress]{School of Computing and Mathematics, Keele University, Staffordshire ST5 5BG, UK}

\begin{abstract}
We derive a one-dimensional (1d) model for the analysis of bulging or necking in an inflated hyperelastic tube of {\it finite wall thickness} from the three-dimensional (3d) finite elasticity theory by applying the dimension reduction methodology proposed by Audoly and Hutchinson (J. Mech. Phys. Solids, \textbf{97}, 2016). The 1d model makes it much easier to characterize fully nonlinear axisymmetric deformations of a thick-walled tube using simple numerical schemes such as the finite difference method. The new model recovers the diffuse interface model for analyzing bulging in a membrane tube and the 1d model for investigating necking in a stretched solid cylinder as two limiting cases. It is consistent with, but significantly refines, the exact linear and weakly nonlinear bifurcation analyses. Comparisons with finite element simulations show that for the bulging problem, the 1d model is capable of describing the entire bulging process accurately, from initiation, growth, to propagation. The 1d model provides a stepping stone from which similar 1d models can be derived and used to study other effects such as anisotropy and electric loading, and other phenomena such as rupture.
\end{abstract}

\end{frontmatter}

\section{Introduction}
Hyperelastic tubes are commonly found in various applications ranging from soft robotics \citep{mh2015, laly2015, lmw2020,stano2021additive} to energy harvesting
\citep{ls2012, bh2013, smith2016,collins2021flexible,bastola2021shape}. They are also used to model human arteries in order to understand pathological conditions such as aneurysms \citep{frz2012, arm2014, dm2015, vd2017, hhp2021}.
Inflation of a hyperelastic tube is one of the few boundary value problems in nonlinear elasticity that have closed-form solutions, and it provides the simplest setup to explain bifurcation, localization, loss of convexity, and \lq\lq two-phase" deformations. Thus, understanding this problem is not only important for applications, but  may also shed light on other more complicated stability and bifurcation problems.

Simple inflation experiments with commercially available latex rubber tubes show that localized bulging is the dominant deformation form. For almost all realistic constitutive models for rubber, the pressure versus volume curve has an up-down-up shape under the condition of fixed resultant axial force \citep{gabook1960}. This led \cite{yin1977} and \cite{ch1984} to analyze the final observable configuration as that corresponding to a \lq\lq two-phase" deformation. The subsequent experimental studies carried out by \cite{kyriakides1990inflation, kyriakides1991initiation}, \cite{pg2006} and \cite{gp2008} have provided a clear picture on how a localized bulge initiates, grows and then propagates under fixed axial force or fixed-ends conditions.

When the membrane assumption is made, the governing equations for tube inflation can be viewed as a finite-dimensional spatial dynamical system that has two conservation laws/integrals \citep{pipkin1968}. This realization enabled \cite{fu2008post} to demonstrate explicitly how a localized solution initiates as a zero-wave-number mode from the uniform deformation and how it evolves into a \lq\lq two-phase" state.
The stability of bulging solutions and their sensitivity to imperfections have been studied under the same framework \citep{pearce2010characterization,fu2010stability, fi2015}. Fresh analytical insight into the case of fixed ends has also been obtained. It is shown that the bifurcation condition for this case corresponds to the axial force reaching a maximum at a fixed pressure  \citep{fi2015}; in other words, as pressure is increased, the critical pressure is the value of pressure at which the axial force reaches a maximum when viewed as a function of the axial stretch. Also, in contrast with the case of fixed axial force where the measured pressure approaches a constant value (the propagation pressure), the measured pressure in the case of fixed ends has an up-down-up shape where the right branch approaches a master curve that is independent of the pre-axial-stretch or the tube length \citep{guo2022localized}.

In some practical applications, however, the tube wall may be of moderate or even large thickness and the membrane model no longer applies. For example, in the context of aneurysm formation, a human artery can be as thick as a quarter of its outer radius \citep{muller2008high}, and fiber-reinforcement also seems to reduce the range of validity of the membrane assumption \citep{wang2018effect}.  Thus, recent studies have begun to consider hyperelastic tubes of finite wall thickness.  \cite{fu2016localized} showed that the associated bifurcation condition for localized bulging corresponds to the vanishing of the Jacobian determinant of the internal pressure and the resultant axial force as functions of the azimuthal stretch on the inner surface and the axial stretch; see also \cite{yu2022analytic} for an alternative derivation. This provides a framework under which additional effects such as rotation \citep{wang2017localized}, double-fiber-reinforcement \citep{wang2018effect}, bi-laying \citep{liu2019prevention, ye2019localized}, torsion \citep{althobaiti2022effect}, and surface tension \citep{ef2021ab, ef2021a, ef2021c, emery2023}
 can be assessed in a systematic manner.  \cite{ye2020weakly} conducted a weakly non-linear analysis and derived the bulging solution explicitly. The analytic predictions were corroborated by numerical simulations \citep{lly2020} and experiments \citep{wang2019experimental}.

For tubes of finite wall thickness, the equations that govern their axisymmetric deformations are coupled nonlinear partial differential equations. Although analytic solutions can be obtained in the near-critical regime using asymptotic methods \citep{ye2020weakly}, the complexity of the governing equations forbids any further analytic attempts to understand the bulging evolution further away from the bifurcation point. The post-bifurcation behavior in the fully nonlinear regime has so far only been investigated by resorting to Abaqus simulations \citep{wang2019experimental, lly2020}. This is not satisfactory since the insight provided by full-scale simulations tends to be limited and there are situations where repeated calculations of the bulging profile are required (e.g. in the assessment of the rupture potential \citep{hhp2021}).

A recent series of studies by Audoly and coworkers has opened the possibility that a 1d reduced model can be derived to describe the fully nonlinear evolution of bulging or necking. In the first of this series, \cite{audoly2016analysis}, the authors derived a 1d model for tensile necking localization in a 3d prismatic solid of arbitrary cross-section. The key idea of their derivation is  a dimension reduction assuming slow variation in the axial direction that respects self-consistency. In terms of the language of perturbation analysis, the leading-order solution is almost correct and higher-order terms are only added to restore self-consistency.
The method was later applied by Lestrigant and Audoly  to obtain a diffuse interface model for the characterization of propagating bulges in membrane tubes \citep{lestringant2018diffuse} and a 1d  model for predicting surface tension-driven necking in soft elastic cylinders \citep{lestringant2020one}. It has also been used recently to derive a 1d model for elastic ribbons \citep{audoly2021one} and for tape springs \citep{kumar2022asymptotic}. The systematic reduction method for deriving 1d strain-gradient models for nonlinear slender structures was further generalized by \cite{lestringant2020asymptotically}. It is worth pointing out that although the 1d models are built on the assumption that localized solutions vary slowly in the longitudinal direction, it is surprisingly accurate, even in the region where the localization is well developed. This is illustrated by the numeric examples in the aforementioned work and in the comparative studies by \cite{wang2021necking} and \cite{fjg2021}.

This work aims to extend the diffuse interface model of \cite{lestringant2018diffuse} for membrane tubes to tubes of finite wall thickness, in a similar spirit as the previous work \cite{fu2016localized} and \cite{ye2020weakly}  that extend the bifurcation condition and the weakly nonlinear analysis from membrane tubes to thick-walled tubes. In contrast with the case under the membrane assumption where the original governing equations are already one-dimensional, the governing equations for the current case are two-dimensional, and the uniformly inflated deformation is no longer homogeneous since the solution depends on the radial variable. It will be shown that a 1d reduced model can still be derived with the associated energy functional simplified to the form
\begin{align}\la{sum}
\mathcal{E}_{\text{1d}}[\muu]=\int_{-L}^L \Big(G(\muu,\lambda(\muu))+\frac{1}{2} D(\muu)\muu'(Z)^2\Big)\,dZ	+C(\muu)\muu'(Z)|_{-L}^L,
\end{align}
where $L$ is the initial half length, $Z$ is the axial coordinate, $a(Z)$ is the azimuthal stretch on the inner surface (a constant multiple of the deformed inner radius ) and the expressions for $G(a,\lambda(a))$, $D(a)$ and $C(a)$ are given in \eqref{eq:G}, \eqref{eq:D} and \eqref{eq:C}, respectively. The first term $G$ in \rr{sum} corresponds to the energy of the uniform deformation, which determines the amplitudes of the two phases in the bulge propagation stage; the second term accounts for the contribution of the strain gradient to the total energy, which describes how the two phases are connected. The Euler-Lagrange equation associated with the energy functional \rr{sum} is a second-order nonlinear ordinary differential equation  for $a(Z)$, which is a drastic simplification from the original nonlinear partial differential equations. This 1d model is validated  by comparison with finite element simulations, showing excellent agreement with numerical results even for the propagation stage.

The outline of this paper is as follows. In Section~\ref{sec:3d}, we formulate the 3d axisymmetric finite-strain model for a tube of finite wall thickness under inflation and axial stretching. In Section~\ref{sec:uniform}, we summarize solutions corresponding to uniform inflation of the tube, making preparation for the subsequent dimension reduction. In Section~\ref{sec:dimension}, we carry out the dimension reduction and derive the aforementioned 1d strain-gradient model. The connection of the 1d  model with prior work is given in Section~\ref{sec:connection}. In Section~\ref{sec:numerical}, we validate the 1d model by making comparison with finite element simulations. Finally, concluding remarks are given in Section~\ref{sec:con}.

\section{Three-dimensional finite-strain  model}\label{sec:3d}
\begin{figure}[h!]
	\centering
	\subfloat[]{\includegraphics[width=0.41\textwidth]{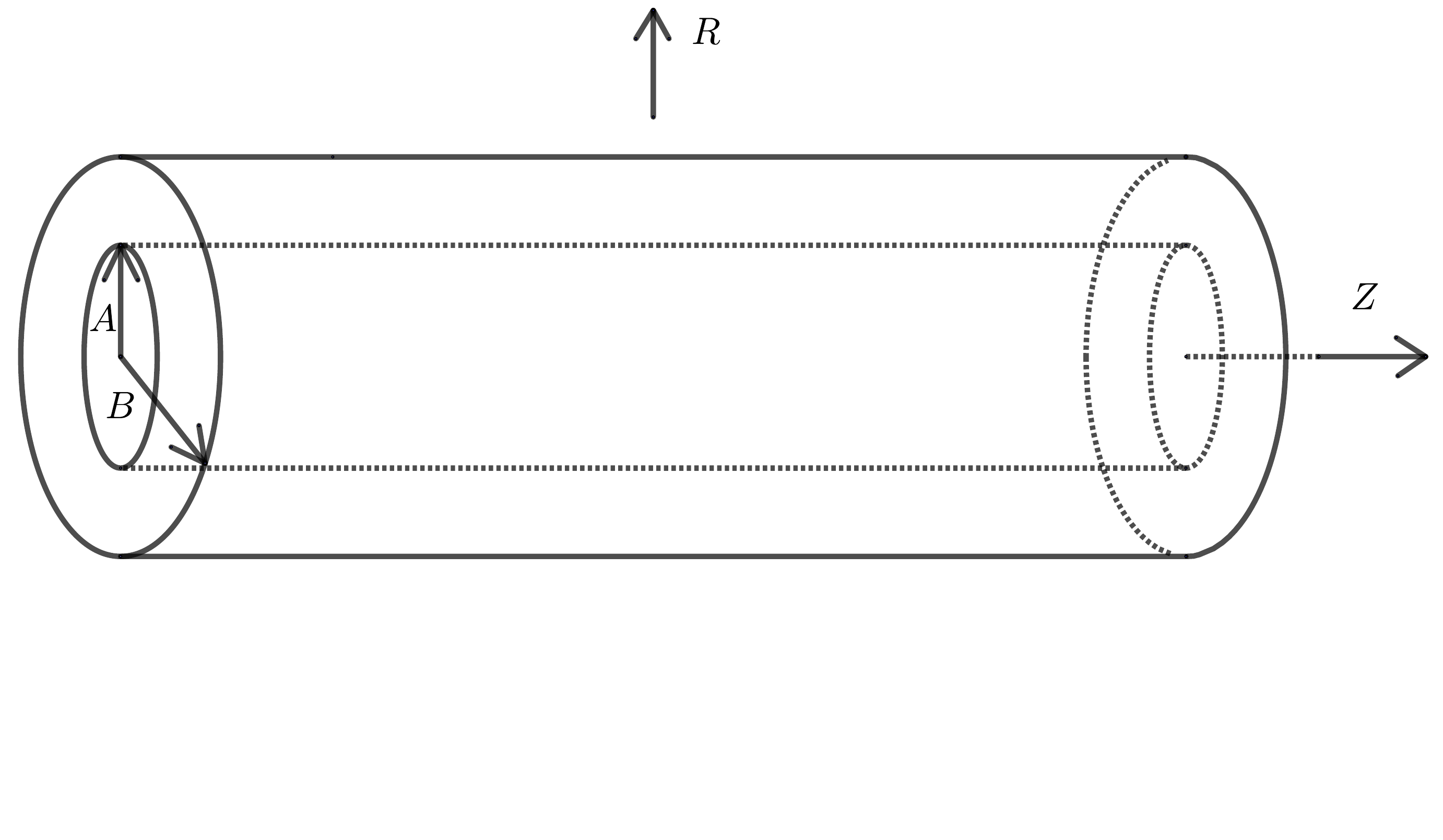}
	}\qquad\quad
	\subfloat[]{\includegraphics[width=0.45\textwidth]{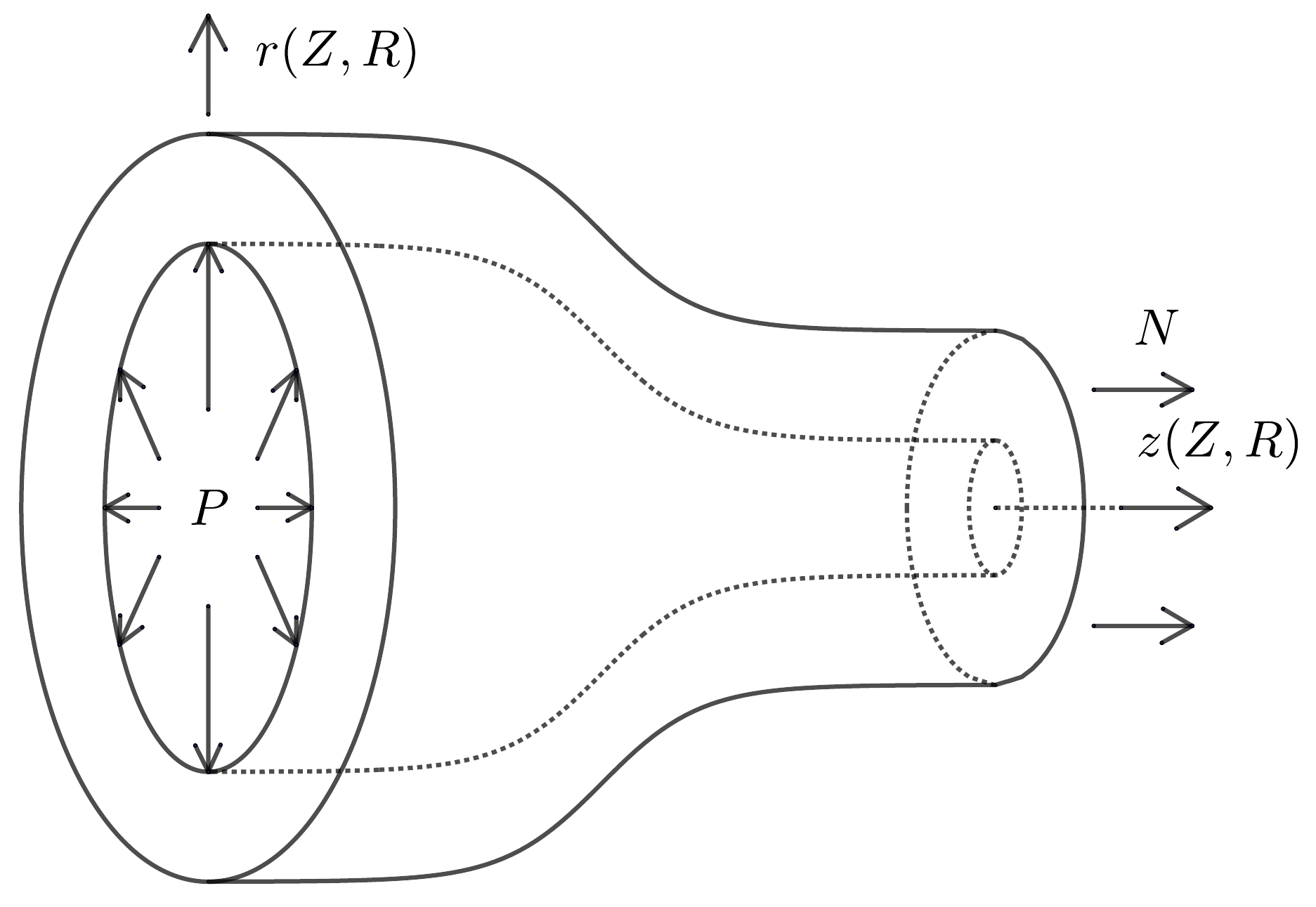}
	}
	\caption{A hyperelastic cylindrical tube of finite wall thickness in (a) reference (undeformed) configuration and (b) current configuration.}
	\label{fig:geometry}
\end{figure}

We consider a circular cylindrical tube that has a length $2L$, inner radius $A$ and outer radius $B$ in its reference configuration; see Fig. \ref{fig:geometry}(a). The ratio of the outer radius to the length $\varepsilon=B/2L$ is assumed to be small; thus $\varepsilon\ll 1$. The tube deforms axisymmetrically under the combined action of an internal pressure $P$ and a resultant axial force $N$, as shown in Fig. \ref{fig:geometry}(b). In terms of cylindrical coordinates,  the current position vector of a representative point is given by
\begin{align}\label{eq:rz}
\bm{x}=z(Z,R)\bm{e}_z+r(Z,R)\bm{e}_r,
\end{align}
where $(R,\Theta,Z)$ and $(r,\theta,z)$ are the coordinates of a representative point before and after deformation, and $(\bm{e}_{r},\bm{e}_\theta,\bm{e}_z)$ are the standard basis vectors associated with both $(R, \Theta, Z)$ and $(r, \theta, z)$.  The deformation gradient related to \eqref{eq:rz} is given by
\begin{align}\label{eq:Ff}
\bm{F}=\frac{r}{R}\bm{e}_{\theta}\otimes \bm{e}_{\theta}+z_Z\bm{e}_{z}\otimes \bm{e}_z+z_R\bm{e}_{z}\otimes \bm{e}_r+r_Z\bm{e}_{r}\otimes \bm{e}_z+r_R\bm{e}_{r}\otimes \bm{e}_r,
\end{align}
where $z_Z:=\partial z/\partial Z$, $z_R:=\partial z/\partial R$, etc.

We assume that the tube is made of an incompressible isotropic hyperelastic material, associated with the strain energy function $W(\lambda_1,\lambda_2,\lambda_3)$, where $\lambda_1$, $\lambda_2$, $\lambda_3$ denote the three principal stretches. Throughout this paper, we identify the indices $1$, $2$, $3$ such that in uniform inflation they coincide with the $\theta$-, $z$- and $r$-directions, respectively.

The total potential energy of the tube is composed of the elastic energy and the load potential, which reads
\begin{align}\label{eq:Ee}
\mathcal{E}=&\int_{-L}^L \Big(\int_A^B \big(w(\lambda_1,\lambda_2)-N^* z_Z\big)2\pi R\,dR-P\pi r^2z_Z\big|_{R=A}\Big)\,dZ,
\end{align}
where $w(\lambda_1,\lambda_2)=W(\lambda_1,\lambda_2,\lambda_1^{-1}\lambda_{2}^{-1})$ is the reduced strain energy function and $N^*=N/(\pi(B^2-A^2))$ is the resultant axial force per unit cross-sectional area. The elastic model governed by the energy functional \eqref{eq:Ee} will be used as a starting point for the subsequent dimension reduction. The governing equations for the two unknown functions $r(Z,R)$ and $z(Z,R)$ can be derived by setting the first variation of $\mathcal{E}$ to zero, but these equations are not required in the approach that we follow.

\section{Uniform inflation}\label{sec:uniform}

We now summarise the solution that corresponds to uniform inflation and extension of the tube. This solution will be referred to as the {\it uniform solution} and is indicated by a superposed bar. For a more detailed derivation, see \cite{ho1979b}.

First, incompressibility implies that the uniform solution must necessarily be of the form
\begin{align}\label{eq:rzbar}
&\bar{z}=\lambda Z,\quad \bar{r}=\sqrt{\muu^2 A^2+\lambda^{-1}(R^2-A^2)},
\end{align}
where $\lambda$  and $\muu$ denote the constant axial stretch and azimuthal stretch on the inner surface, respectively.
The three principal stretches are simply
\begin{align}\label{eq:lambda}
\bar{\lambda}_1=\frac{\bar{r}}{R},\quad \bar{\lambda}_2=\lambda,\quad \bar{\lambda}_3=\frac{d \bar{r}}{dR}=\bar{\lambda}_1^{-1}\bar{\lambda}_2^{-1},
\end{align}
and the azimuthal stretch on the outer surface, denoted by $\muub$, is given by
\begin{align}
\muub=\bar{\lambda}_1|_{R=B}=\frac{{\sqrt{\muu^2 A^2+\lambda^{-1}(B^2-A^2)}}}{B}.
\end{align}
The three associated principal Cauchy stresses $\bar{\sigma}_{11}, \bar{\sigma}_{22}$ and $\bar{\sigma}_{33}$ satisfy the relations
\begin{align}
\bar{\sigma}_{11}- \bar{\sigma}_{33}=\bar{\lambda}_1 w_1,\quad \bar{\sigma}_{22}- \bar{\sigma}_{33}={\lambda} w_2, \la{stresses}
\end{align}
where $w_1=\paa w(\bar{\lambda}_1, \bar{\lambda}_2)/\paa \bar{\lambda}_1$ and $w_2=\paa w(\bar{\lambda}_1, \bar{\lambda}_2)/\paa \bar{\lambda}_2$.

The only equilibrium equation that is not satisfied automatically is
\be
\frac{d \bar{\sigma}_{33}}{d \bar{r}}=\frac{\bar{\sigma}_{11}-\bar{\sigma}_{33}}{\bar{r}}=\frac{\bar{\lambda}_1 w_1}{\bar{r}}. \la{sig33} \en
On integrating this equation from $R=A$ to $R=B$ and making use of the boundary conditions that $\bar{\sigma}_{33}|_{R=A}=-P$ and $\bar{\sigma}_{33}|_{R=B}=0$, we obtain
\begin{align}
P=Q(\muu,\lambda):= \int_{\muub}^{\muu}\frac{w_1(\bar{\lambda}_1,\lambda)}{\bar{\lambda}_1^2\lambda-1}\,d\bar{\lambda}_1,\label{eq:eqQ}
\end{align}
where the second equation defines the function $Q(\muu,\lambda)$ and we have made  use  of the identity
\begin{align}
\frac{d\bar{r}}{\bar{r}}=-\frac{d\bar{\lambda}_1}{\bar{\lambda}_1(\bar{\lambda}_1^2\lambda-1)},
\end{align}
which can be deduced from \rr{eq:rzbar}$_2$ and \rr{eq:lambda}$_1$.

The overall equilibrium in the axial direction implies
\begin{align}\label{eq:M}
M(\muu,\lambda)-\frac{1}{2}  \muu^2P  -\frac{N}{2\pi A^2}=0,
\end{align}
where $M(\muu,\lambda)$ is given by
\begin{align}
M(\muu,\lambda):= \frac{1}{ A^2} \int^{B}_{A} \lambda^{-1} \bar{\sigma}_{22} R\, dR
  =\int_{\muub}^\muu \frac{
	(\bar{\lambda}_1^2-\muu^2)w_1(\bar{\lambda}_1,\lambda)+2 \bar{\lambda}_1  \lambda ( \muu^2\lambda-1) w_2(\bar{\lambda}_1,\lambda) }{2(\bar{\lambda}_1^2\lambda-1)^2}\,d\bar{\lambda}_1.\label{eq:eqM} \
\end{align}

In view of \eqref{eq:Ee}, the total potential energy  of the uniform deformation \eqref{eq:rzbar} per unit reference length, after scaling by $2\pi$,  is
\begin{align}\label{eq:G}
G(\muu,\lambda)= \int_A^B w(\bar{\lambda}_1,\lambda)\, R\,dR- \frac{1}{2} P A^2\muu^2\lambda-\frac{N}{2 \pi}\lambda.
\end{align}
The equilibrium equations \rr{eq:eqQ} and \rr{eq:M} can also be obtained from  $\partial G/\partial\muu=0$ and $\partial G/\partial \lambda=0$, respectively.
Once the loads $P$ and $N$ are specified, the deformation parameters $\muu$ and $\lambda$ can be found by solving the equilibrium equations \eqref{eq:eqQ} and \eqref{eq:M}.

On differentiating the left-hand side of \eqref{eq:M} with respect to $\lambda$, we find that its derivative takes the form 
$H w_{22}(\muu,\lambda)/A+O(H^2)$, where $H=B-A$ is the thickness of the tube. We assume that the strong ellipticity condition is satisfied pointwise which guarantees that $w_{22}(\muu,\lambda)$ is positive \citep{ks1976}. This, combined with the implicit function theorem, implies that \eqref{eq:M} can be inverted to express $\lambda$ in terms of $\muu$ uniquely at least when $H$ is small. We assume that this remains true for arbitrary $H$. This enables us to view \eqref{eq:M} as an implicit equation for $\lambda=\lambda(\muu)$. We remark that $\lambda$ is also dependent on $P$, but this dependence is not indicated explicitly for notational brevity. Thus, by definition, $\lambda(\muu)$ is the solution to the implicit equation
\begin{align}\label{eq:lambdamu}
M(\muu,\lambda(\muu))-\frac{1}{2}\muu^2 P -\frac{N}{2\pi A^2}=0.
\end{align}
Since $\lambda$ has been viewed as a  function of $\muu$, all quantities (except $\bar{z}$ which also depends on $Z$) related to the uniform solution  are functions of $\muu$ and $R$. For instance, $\bar{r}$ now denotes the function
\begin{align}\label{eq:rr}
\bar{r}(\muu,R)=\sqrt{\muu^2 A^2+\lambda(\muu)^{-1}(R^2-A^2)}.
\end{align}
We denote  $\bar{\sigma}_{33}$ by $-q(\muu, R)$ so that
\begin{align}
q(\muu, R):= -\bar{\sigma}_{33}= \int^{\bar{\lambda}_1}_{\muub}\frac{w_1(\tilde{\lambda}_1,\lambda)}{\tilde{\lambda}_1^2\lambda-1}\,d\tilde{\lambda}_1.\label{eq:eqQq}
\end{align}
We also define another function $m(\muu,R)$ through
\begin{align}
m(\muu,R):=\frac{1}{ R^2} \int^{B}_{R} \lambda^{-1} \bar{\sigma}_{22} T\, dT =\int_{\muub}^{\bar{\lambda}_1} \frac{(\tilde{\lambda}_1^2-\bar{\lambda}_1^2)w_1(\tilde{\lambda}_1,\lambda)+2  \tilde{\lambda}_1 \lambda ( \bar{\lambda}_1^2\lambda-1) w_2 (\tilde{\lambda}_1,\lambda)}{2(\tilde{\lambda}_1^2\lambda-1)^2}\,d\tilde{\lambda}_1, \label{eq:m}
\end{align}
and record the connections
\begin{align}
q(\muu, A)=Q(\muu, \lambda(\muu)),\quad m(\muu, A)=M(a, \lambda(a)). \label{eq:q1}
\end{align}
The 1d reduced model to be derived in the next section will be expressed in terms of the two functions $q(\muu, R)$ and $m(\muu, R)$. The integrals in these two functions can be evaluated explicitly for some commonly used strain energy functions, including the neo-Hookean, Mooney-Rivlin and Gent material models. The last one will be used in our illustrative examples.
\section{Derivation of the one-dimensional model}\label{sec:dimension}

In this section, we apply the dimension reduction methodology proposed by \cite{audoly2016analysis} to derive a one-dimensional model from the full three-dimensional model formulated in Section~\ref{sec:3d}.

\subsection{Optimal correction}

 We start our dimension reduction by assuming that all dependent variables related to the axisymmetric configuration vary slowly in the axial direction. More precisely, it is assumed that all variables depend on $Z$ through the \lq\lq far distance" variable
\begin{align}
S=\varepsilon Z.
\end{align}
Recall that $\varepsilon$ is the ratio of the outer radius to the length, which is assumed to be small. In particular, we now allow $\muu$ and $\lambda$ to depend on $S$ and write $\muu=\muu(S)$, $\lambda=\lambda(\muu(S))$. Our aim is to derive a reduced model, an ordinary differential equation, that is satisfied by $\muu(S)$. We recall that $\muu(S)$ is the deformed inner radius divided by a constant (i.e. $A$).

A naive approach would be to use $\muu=\muu(S)$ and $\lambda=\lambda(\muu(S))$ to compute the two principal stretches and then derive the equation satisfied by $\muu=\muu(S)$ by minimizing the energy functional \rr{eq:Ee}. However, this would yield an equation for $a(S)$ that is not self-consistent. The correct way is to allow for higher-order correction terms by looking for an asymptotic solution of the form
\begin{align}\label{eq:zrp}
\begin{split}
&z(Z,R)=\frac{1}{\varepsilon}\int_0^{S} \lambda(\muu(T))\,dT+\varepsilon v^*(S,R)+O(\varepsilon^3),\\
&r(Z,R)=\bar{r}(\muu(S),R)+\varepsilon^2 u^*(S,R)+O(\varepsilon^4),
\end{split}
\end{align}
where $u^*$ satisfies the kinematic constraint 
\begin{align}\label{eq:kinematic}
u^*(S,A)=0,
\end{align}
ensuring that $a$ represents the azimuthal stretch on the inner surface. We note that the correction terms in $z(Z,R)$ and $r(Z,R)$ are of order $\varepsilon$ and $\varepsilon^2$, respectively. This is because the $O(1)$-term in $z(Z,R)$ and the $O(\varepsilon)$-term in $r(Z,R)$  correspond to a uniform perturbation and  can thus be absorbed into the leading-order terms.

On substituting \eqref{eq:zrp} into \eqref{eq:Ff} and truncating at order $\varepsilon^2$, we obtain the deformation gradient
\begin{align}
\bm{F}=\begin{pmatrix}
{\bar{r}}/{R}+\varepsilon^2 {u^*}/{R} & 0 & 0 \\
0 & \lambda(\muu(S))+\varepsilon^2 v^*_S & \varepsilon v^*_R\\
0 & \varepsilon \bar{r}_a \muu'(S)  & \bar{r}_R+\varepsilon^2 u^*_R
\end{pmatrix},
\end{align}
where the subscripts represent partial differentiation with respect to the indicated variables (in particular $\bar{r}_a=\paa \bar{r}/\paa a$). Consequently, the principal stretches $\lambda_1$ and $\lambda_2$ are given by
\begin{align}
\begin{split}\label{eq:lam12}
&\lambda_1=\bar{\lambda}_1+\varepsilon^2 \frac{u^*}{R}, \\
&\lambda_2=\bar{\lambda}_2+\varepsilon^2\Big(v^*_S+  \frac{\lambda ( \bar{r}_a^2 \muu'(S)^2+v^{*2}_R)+2\bar{\lambda}_3 \bar{r}_a \muu'(S)   v^*_R }{2(\lambda^2-\bar{\lambda}_3)}\Big),
\end{split}
\end{align}
where $\bar{\lambda}_1$, $\bar{\lambda}_2$ and $\bar{\lambda}_3$  are still given by \eqref{eq:lambda} but with $\muu$ and $\lambda$ replaced by $\muu(S)$  and $\lambda(\muu(S))$, respectively.

Substituting \eqref{eq:lam12} into \eqref{eq:Ee} and expanding to order $\varepsilon^2$,  we see that $\mathcal{E}$ can be written, in terms of the un-scaled variables, as
\begin{align}\label{eq:te}
\mathcal{E}=2\pi \Big(\int_{-L}^L  G(\muu(Z),\lambda(\muu(Z)))\,dZ+\mathcal{E}_2\Big)+O(L \varepsilon^3),
\end{align}
where $\mathcal{E}_2$ represents the term of order $\varepsilon^2$ and is given by
\begin{align}\label{eq:e2}
\begin{split}
\mathcal{E}_2=& \int_{-L}^L\Big( \int_{A}^B \Big( (w_2-N^*) v_Z+  w_2\frac{\lambda (\bar{r}_a^2 \muu'^2+v_R^2)+2 \bar{\lambda}_3  \bar{r}_a \muu'  v_R }{2(\lambda^2-\bar{\lambda}_3^2)}\Big)R\,dR\\
&+\int_A^B  w_1u \,dR-\frac{1}{2}PA^2 \muu^2 v_Z|_{R=A}\Big)\,dZ.	
\end{split}
\end{align}
In the above expression, $v(Z,R)=\varepsilon v^*(S,R)$ and $u(Z,R)=\varepsilon^2 u^*(S,R)$ denote the unscaled displacements, and here and hereafter we write $a(Z)$ for $a(S)$ and so $a'$ now denotes $a'(Z)$. It is seen that the only reason for introducing $S$ above is to identify all terms of order $\varepsilon^2$ that should be kept in \rr{eq:e2}. With this task accomplished,
the scaled variable $S$ will no longer appear in the subsequent analysis. Also,
$w_1=w_1(\bar{\lambda}_1,\lambda)$,  $w_2=w_2(\bar{\lambda}_1,\lambda)$  in which $\lambda$ is a function of $a$ and $\bar{\lambda}_1$ is a function of $a$ and $R$.

Our formulation in terms of the reduced strain energy function requires the solution \rr{eq:zrp} to satisfy the incompressibility condition automatically. This can be achieved by eliminating $u$ in \eqref{eq:e2} with the use of $\det(\bm{F})=1$ which takes the form
\begin{align}\label{eq:incon}
\lambda ( \bar{r} u)_R +\bar{r}(\bar{r}_R v_Z-\bar{r}_a \muu' v_R )=0.
\end{align}
To this end, we make use of the relation $d\bar{\sigma}_{33}/{dR}={w_1}/{(\lambda \bar{r})}$ which follows from \eqref{eq:eqQq} and write
\begin{align}\label{eq:w1u}
\begin{split}
\int_A^B w_1 u\, dR&=\lambda\int_A^B \bar{\sigma}_{33,R} \bar{r}  u\,dR=\lambda \bar{\sigma}_{33}  \bar{r}  u|_{A}^{B}-\lambda \int_{A}^B  \bar{\sigma}_{33} (\bar{r}u)_R\, dR\\
&=-  \int_{A}^B  q \bar{r}(\bar{r}_R v_Z-\bar{r}_a \muu' v_R )\, dR,
\end{split}
\end{align}
where the boundary term $\lambda \bar{\sigma}_{33}  \bar{r}  u|_{A}^{B}$ vanishes because $\bar{\sigma}_{33}|_{B}=0$ and the kinematic condition \eqref{eq:kinematic} implies $u|_{A}=0$; furthermore, we have replaced $\bar{\sigma}_{33}$ by $-q(\muu, R)$ (cf. \rr{eq:eqQq}) and have used \rr{eq:incon} to eliminate $( \bar{r} u)_R$.

On eliminating $u$ in \eqref{eq:e2} with the use of \eqref{eq:w1u}, we can recast $\mathcal{E}_2$ in the form
\begin{align}\label{eq:e2b}
\begin{split}
\mathcal{E}_2=& \int_{-L}^L\Big( \int_{A}^B \big((\lambda^{-1}\bar{\sigma}_{22}-N^*) v_Z+  \frac{1}{2}\zeta (\bar{r}_a^2 \muu'^2+v_R^2)+\xi \bar{r}_a \muu' v_R\big)R\,dR\\
&-\frac{1}{2}PA^2 \muu^2 v_Z|_{R=A}\Big)\,dZ,	
\end{split}
\end{align}
where we have made use of the connection $\lambda w_2-q = \bar{\sigma}_{22}$ that follows from \rr{stresses}$_2$ with $\bar{\sigma}_{33}=-q$, and $\zeta$ and $\xi$ are given by
\begin{align}\label{eq:zetaadd}
\zeta=\frac{\lambda w_2}{\lambda^2-\bar{\lambda}_3^2}, \quad \xi=\frac{\bar{\lambda}_3}{{\lambda}}\zeta+q\bar{\lambda}_1.
\end{align}
 Then upon using integration by parts, we obtain
\begin{align}\label{eq:e2c}
\begin{split}
\mathcal{E}_2=& \int_{-L}^L\Big( \int_{A}^B \big(-(\lambda^{-1}\bar{\sigma}_{22})_a \muu' R v+\frac{1}{2}R \zeta (\bar{r}_a^2 \muu'^2+v_R^2)+R\xi\bar{r}_a \muu'v_R\big)\,dR+PA^2\muu \muu' v|_{R=A}\Big)\,dZ\\
&+\Big(\int_A^B (\lambda^{-1}\bar{\sigma}_{22}-N^*)vR\,dR-\frac{1}{2}PA^2 \muu^2v|_{R=A}\Big)\Big|_{Z=-L}^{Z=L},
\end{split}
\end{align}
where $(\lambda^{-1}\bar{\sigma}_{22})_a$ denotes the partial derivative of $\lambda^{-1}\bar{\sigma}_{22}$ with respect to $a$ with $R$ fixed. A repeated application of integration by parts using \eqref{eq:m} allows us to write the first integral in \eqref{eq:e2c} as
\begin{align}\label{eq:ma}
\begin{split}
\int_A^B-(\lambda^{-1}\bar{\sigma}_{22})_a \muu' R v\,dR&=\int_A^B (R^2m(a,R))_{a,R}a'v\,dR\\
&=A^2\frac{\partial}{\partial a}m(a,A) a'v|_{R=A}-\int_A^B R^2\frac{\partial}{\partial a}m(a,R)a'v_R\,dR,
\end{split}
\end{align}
where we have used the relation $m(a,B)=0$ to eliminate the boundary term at $R=B$. Upon substituting \eqref{eq:ma} into \eqref{eq:e2c}, the second-order energy $\mathcal{E}_2$ can be rewritten as
\begin{align}
\begin{split}\label{eq:e2d}
\mathcal{E}_2=& \int_{-L}^L\Big(\int_A^B \big(\frac{1}{2}R \zeta (\bar{r}_a^2 \muu'^2+v_R^2)+R\xi\bar{r}_a \muu'v_R- R^2\frac{\partial}{\partial a}m(a,R)a'v_R\big) \,dR\\
&+\Big(\int_A^B (\lambda^{-1}\bar{\sigma}_{22}-N^*)vR\,dR-\frac{1}{2}PA^2 \muu^2v|_{R=A}\Big)\Big|_{Z=-L}^{Z=L},
\end{split}
\end{align}
where we have used the equality ${\partial m(a,A)}/{\partial a}=P\muu$ implied by \eqref{eq:lambdamu}  and $\eqref{eq:q1}_2$  to simplify $\mathcal{E}_2$.

To find the remaining correction field $v=v(Z,R)$, we treat the leading-order stretch $a(Z)$ as stipulated and seek the correction $v$ such that the total
potential energy is stationary \citep{audoly2016analysis}. By completing the square technique, we can write the first integrand in \eqref{eq:e2d} as 
\begin{align}
\begin{split}
&\frac{1}{2}R \zeta (\bar{r}_a^2 \muu'^2+v_R^2)+R\xi\bar{r}_a \muu'v_R- R^2\frac{\partial}{\partial a}m(a,R) a'v_R\\
=&\frac{1}{2}R\zeta(v_R-c(a,R)a')^2+\frac{1}{2}R\zeta  a'^2(\bar{r}_a^2-c(a,R)^2),
\end{split}
\end{align}
where $c(\muu,R)$ is defined by
\begin{align}\label{eq:c}
c(\muu,R)&=-\frac{\bar{r}_a}{\bar{\lambda}_1 \lambda^2 }+\frac{1}{ R \zeta}\Big(R^2 \frac{\partial}{\partial\muu}m(\muu,R)- \bar{r}\bar{r}_a q(\muu,R)\Big),
\end{align}
Note that we have used \eqref{eq:m} and \eqref{eq:zetaadd} to achieve the simple form \eqref{eq:c}. Consequently, $\mathcal{E}_2$ and the potential energy \eqref{eq:te} are minimized when 
\begin{align}\label{eq:vR}
v_R=c(a,R)a'(Z).
\end{align}
Once $v_R$ is found, the optimal correction $v$ can be obtained by  integrating \eqref{eq:vR} from $B$ to $R$, which yields
\begin{align}\label{eq:v}
v=-\left(\int_{R}^B c(\muu,T)\,dT\right)\muu'(Z),
\end{align}
where  we have neglected the function arising from integration since it  does not enter the  potential energy.

\subsection{Energy functional corresponding to the 1d reduced model}

Substituting the correction function $v$ found in \eqref{eq:v} back into \eqref{eq:e2d}, after some simplification, we obtain the final expression for the energy functional of the 1d model
\begin{align}\label{eq:1d}
\mathcal{E}_{\text{1d}}[\muu]=\int_{-L}^L \Big(G(\muu,\lambda(\muu))+\frac{1}{2} D(\muu)\muu'(Z)^2\Big)\,dZ	+C(\muu)\muu'(Z)|_{-L}^L,
\end{align}
where the gradient moduli $D$ and $C$ are given by
\begin{align}
&D(\muu)= \int_A^B R\zeta (\bar{r}_a^2-c(\muu,R)^2)\,dR,\label{eq:D}\\	
\begin{split}
&C(\muu)=\int_A^B (\lambda^{-1}\bar{\sigma}_{22}-N^*)(\tilde{c}(\muu,R)-\tilde{c}(\muu,A))R\,dR,\label{eq:C}
\end{split}
\end{align}
with $\tilde{c}(\muu,R)=-\int_{R}^B c(\muu,T)\,dT$.

The associated equilibrium equation is obtained by extremizing \eqref{eq:1d} with respect to $\muu(Z)$ and is found to take the form
\begin{align}\label{eq:el}
A^2\muu\lambda(\muu)(Q(\muu,\lambda(\muu))-P)-\frac{1}{2}D'(\muu)\muu'(Z)^2-D(\muu)\muu''(Z)=0,
\end{align}
where we have used the fact that $\partial G/\partial \lambda=0$  as it is used to find the implicit relation between $\lambda$ and $\muu$ (see \eqref{eq:lambdamu}).
Since $Z$ does not explicitly  appear in the integrand of \eqref{eq:1d} due to the translational invariance of the current problem in $Z$, by the Beltrami identity, the equilibrium equation \eqref{eq:el} admits a first integral of the form
\begin{align}\label{eq:first}
G(\muu,\lambda(\muu))-\frac{1}{2}D(\muu)\muu'(Z)^2=\text{constant}.
\end{align}

We remark that the variational problem \eqref{eq:1d} is ill-posed due to the presence of the boundary terms $C(\muu)\muu'(Z)|_{-L}^{L}$. This is because the variational structure of the problem is broken when higher-order terms are dropped. There are two possible ways to get around this issue \citep{lestringant2020asymptotically}. The first one is to simply ignore the boundary terms, i.e., to set $C(\muu)=0$. The second one is to add an $O(\varepsilon^2)$-term to $\muu(Z)$ so that the boundary terms go away, which  is rigorous but slightly more complex. It has previously been verified in \cite{lestringant2020asymptotically} that the simple and rigorous approaches yield curves that can hardly be distinguished visually in any of the plots.

To summarize, the second-order nonlinear ordinary differential equation \rr{eq:el} is our approximate 1d model that governs the variation of the inner radius (which is $A$ times $a(Z)$) in the axial direction. Once $a(Z)$ is determined, the 3d deformation is given by \rr{eq:rzbar}. We note that the function $Q(\muu,\lambda(\muu))$ is explicit for most of the commonly used strain energy functions. The only slight complication is that the function $D(a)$ is given by an integral; see \rr{eq:D}, but the functions $m(a, R)$, $q(a, R)$, and hence $c(a, R)$ and the integrand in \rr{eq:D} all have explicit expressions for most of the commonly used strain energy functions. Thus, only one numerical integration is required. This can easily be implemented on a symbolic manipulation platform such as {\it Mathematica} \citep{wo1991} as we shall show later.

\section{Connections with previous work}\label{sec:connection}

We now demonstrate that the 1d model derived in Section~\ref{sec:dimension} can recover the 1d model of \cite{lestringant2018diffuse} for membrane tubes and that of  \cite{audoly2016analysis} for solid cylinders under appropriate limits, and it can also reproduce the same weakly nonlinear bulging solution as that based on the exact 3d theory \citep{ye2020weakly}.

\subsection{Membrane limit}
We first consider the reduction of the 1d model in the membrane limit where the tube thickness $H$ approaches zero.
The general axisymmetric deformation of a membrane tube is described by
\begin{align}
r=r(Z), \quad \theta=\Theta, \;\;\;\; z=z(Z),
\end{align}
and the three principal stretches are given by
\begin{align}
\lambda_1=\frac{r}{R}, \quad \lambda_2=\sqrt{r'^2+z'^2},\quad \lambda_3=1/(\lambda_1 \lambda_2),
\end{align}
where $R$ denotes the constant radius of the mid-surface.
The total energy \rr{eq:Ee} reduces to
\begin{align}
\mathcal{E}= 2 \pi  \int_{-L}^{L} \Big(w-\frac{1}{2} {P^*} \lambda_1^2 z'-{N^*} z'\Big)\, dZ, \la{lll}
\end{align}
where $P^*$ denotes the pressure scaled by $H/R$.
Setting the first variation $\delta \mathcal{E}$ to be zero then gives the governing equations
\begin{align}
&w_1 - R \Big(\frac{w_2}{\lambda_2} r'\Big)' -P^* \lambda_1 z'=0, \\
&\frac{w_2}{\lambda_2} z'-\frac{1}{2} P^* \lambda_1^2   =N^*.\la{memeq}
\end{align}
 Under the assumption that $|r'| \ll 1$, we have
\be \lambda_2=z' +\frac{r'^2}{2 z'}+\cdots. \la{lam2} \en
As an algebraic equation for $z'$, Eq. \rr{memeq} has an asymptotic solution of the form
\begin{align}
 z'=g(\lambda_1)+k_1(\lambda_1)r'^2+\cdots, \la{zd}
\end{align}
where the leading-order term $g(\lambda_1)$ obviously satisfies the algebraic equation
\begin{align}
w_2(\lambda_1, g(\lambda_1)) -\frac{1}{2} P^* \lambda_1^2 -N^*=0, \la{gr}
\end{align}
and the function $k_1(\lambda_1)$ can easily be found but is not required. Eq. \rr{gr} determines $g(\lambda_1)$ uniquely under the assumption $w_{22}>0$.

With the use of \rr{lam2} and \rr{zd}, we may expand the integrand in \rr{lll} to order $r'^2$ and obtain
\begin{align}
\mathcal{E}= 2 \pi \int_{-L}^{L} \Big( w(\lambda_1, g(\lambda_1))-\frac{1}{2}  P^* \lambda_1^2 g(\lambda_1)- N^* g(\lambda_1)+\frac{1}{2}\frac{w_2(\lambda_1, g(\lambda_1))}{ g(\lambda_1)} r'^2 \Big)\, dZ. \la{mem0}
\end{align}
This is the reduced model derived by \cite{lestringant2018diffuse}.

We now show that our general 1d model \rr{eq:1d} can recover this 1d model in the limit $H \to 0$. To this end, we first note that the
 uniformly deformed configuration in the zero-thickness limit is given by
\begin{align}
\bar{r}=\muu R, \quad \bar{z}=\lambda Z.
\end{align}
In particular,  we have $\bar{r}_a=R$. Since $q(\muu,R)$ and $m(\muu,R)$ involve integrals from $R$ to $B$, they go to zero as $H \to 0$. Consequently, the $c(\muu,R)$ defined in \eqref{eq:c} takes the simple form
\begin{align}\label{eq:ff}
c(\muu,R)=-\frac{R}{\muu \lambda^2}.
\end{align}
Taking the limit $H \to 0$ in  \eqref{eq:zetaadd} yields
\begin{align}\label{eq:zeta}
\zeta=\frac{ \muu^2\lambda^3 w_2}{\muu^2\lambda^4-1}.
\end{align}
Substituting \eqref{eq:ff} and \eqref{eq:zeta} into \eqref{eq:1d}, we obtain
\be
\lim_{H\to 0}\frac{\mathcal{E}_{\text{1d}}[\muu]}{H}=R \int_{-L}^L\Big( w(\muu,\lambda(\muu))-\frac{1}{2}P^* \muu^2\lambda(\muu)-N^*\lambda(\muu)
+\frac{1}{2}R^2 \frac{w_2(\muu,\lambda(\muu))}{\lambda(\muu)}\muu'(Z)^2\Big)\,dZ.\la{mem} \en
Note that the modulus $C(\muu)$ vanishes in the membrane limit because it is of order $H^2$.
The integrand on the right-hand side of \rr{mem} is the same as that on the right-hand side of \rr{mem0} if we identify $\lambda_1$, $g(\lambda_1)$ and $r'$ with
$a(Z)$, $\lambda(a)$, and $R a'(Z)$, respectively.

\subsection{Solid cylinder limit}
Next we consider the other extreme limit corresponding to $A\to 0$ and $P\to 0$. in this case, the uniform solution takes the form
\begin{align}\label{eq:uni}
\bar{z}=\lambda Z,\quad \bar{r}=\muu R
\end{align}
with $\muu=\lambda^{-1/2}$. The three principal stretches are
\begin{align}
\bar{\lambda}_1=\bar{\lambda}_3=\lambda^{-1/2},\quad \bar{\lambda}_2=\lambda.
\end{align}
In particular, we have
\begin{align}\label{eq:w1w2}
w_1(\bar{\lambda}_1,\bar{\lambda}_2)=0,\quad w_2(\bar{\lambda}_1,\bar{\lambda}_2)=\hat{w}'(\lambda),
\end{align}
where $\hat{w}(\lambda)=W(\lambda^{-1/2},\lambda,\lambda^{-1/2})$. It follows from  $\eqref{eq:w1w2}_1$ that $q(\muu,R)=0$. Note that the deformation \eqref{eq:uni} is homogeneous, so \eqref{eq:m} implies that
\begin{align*}
m(\muu,R)=\frac{A^2(B^2-R^2)}{R^2(B^2-A^2)}m(\muu,A)=\frac{A^2(B^2-R^2)}{R^2(B^2-A^2)}M(\muu, \lambda(a)).
\end{align*}
Differentiating this expression with respect to $\muu$ and noting  \eqref{eq:lambdamu}, we obtain $\partial m(\muu,R)/\partial\muu=0$. Thus $c(\muu,R)$ reduces to
\begin{align}\label{eq:fA0}
c(\muu,R)=-\frac{R}{\lambda^{3/2} }.
\end{align}
According to \eqref{eq:zetaadd}, the elastic modulus $\zeta$ is easily calculated as
\begin{align}\label{eq:zetaA0}
\zeta=\frac{\lambda^2 \hat{w}'(\lambda)}{\lambda^3-1}.
\end{align}
Substituting \eqref{eq:fA0} and \eqref{eq:zetaA0} into \eqref{eq:1d}, we obtain
\begin{align}
2\pi\mathcal{E}_{\text{1d}}[\lambda]=\int_{-L}^L \Big(\pi B^2 \hat{w}(\lambda)+\frac{\pi B^4 }{16} \frac{\hat{w}'(\lambda)}{\lambda^4}\lambda'(Z)^2-N\lambda\Big)\,dZ,
\end{align}
where we have made use of the relation $\muu'(Z)=-{\lambda'(Z)}/{(2\lambda^{3/2})}$. This recovers the 1d model of \cite{audoly2016analysis} specialized to an incompressible circular cylinder.

\subsection{Comparison with exact weakly nonlinear analysis}\label{subsec:weakly}
Finally, we carry out a weakly nonlinear near-critical analysis using our 1d model and compare the resulting amplitude equation with that obtained by \cite{ye2020weakly} from the exact 3d theory. We focus on localized solutions in an infinitely long tube of finite wall thickness.

Denote by $\muu_\infty$ the limit of $\muu(Z)$ as $Z\to\infty$ and $\lambda_\infty=\lambda(\muu_\infty)$. It follows from \eqref{eq:eqQ} and \eqref{eq:lambdamu}  that
\begin{align}
&P=Q(\muu_\infty,\lambda_\infty),\quad   N=2\pi A^2 F(\muu_\infty,\lambda_\infty)\label{eq:QQMM},
\end{align}
where $F(\muu_\infty,\lambda_\infty)$ is defined by
\begin{align}
F(\muu_\infty,\lambda_\infty)=M(\muu_\infty,\lambda_\infty)-\frac{1}{2}\muu_\infty^2 Q(\muu_\infty,\lambda_\infty).
\end{align}
We look for a localized solution that bifurcates from the uniform solution by writing
\begin{align}\label{eq:muZ}
\muu(Z)=\muu_\infty+y(Z),
\end{align}
where $y(Z)$ is a small perturbation. Substituting  \eqref{eq:muZ} into the 1d equilibrium equation \eqref{eq:el} and expanding in terms of $y(Z)$ to quadratic order with the use of \eqref{eq:QQMM}, we obtain
\begin{align}\label{eq:amp}
D(\muu_\infty)y''(Z)=\omega(\muu_\infty,\lambda_\infty)y(Z)+\gamma(\muu_\infty,\lambda_\infty)y(Z)^2,
\end{align}
where the two coefficient functions $\omega(\muu,\lambda)$ and $\gamma(\muu,\lambda)$ are given by
\begin{align}
&\omega(\muu,\lambda)=A^2\frac{ 2 \muu\lambda}{\muu^2 Q_{\lambda}+2F_{\lambda}}\Omega(\muu,\lambda),\label{eq:omega}\\
&\gamma(\muu,\lambda)=A^2\frac{\muu\lambda(\muu^2 Q_{\muu}+2 F_{\muu})}{ F_{\muu}(\muu^2 Q_{\lambda}+2  F_{\lambda})^2}\Gamma(\muu,\lambda)+A^2\psi(\muu,\lambda)\Omega(\muu,\lambda).\label{eq:omega1}
\end{align}
In the above expressions, $Q_{a}=\partial Q(a,\lambda)/\partial a$, $Q_{\lambda}=\partial Q(a,\lambda)/\partial \lambda$, etc., $\Omega(\muu,\lambda)$ and $\Gamma(\muu,\lambda)$ are defined by
\be
\Omega(\muu,\lambda)=\frac{\partial Q}{\partial\muu}\frac{\partial F}{\partial\lambda}-\frac{\partial Q}{\partial\lambda}\frac{\partial F}{\partial\muu}, \;\;\;\;
\Gamma(\muu,\lambda)=\frac{\partial \Omega}{\partial\muu}\frac{\partial F}{\partial\lambda}-\frac{\partial \Omega}{\partial\lambda}\frac{\partial F}{\partial\muu},\la{addd}
\en
and $\psi(\muu,\lambda)$ is not written out as it is not required  in the weakly nonlinear analysis.

The solution to the linearized equation of \eqref{eq:amp} changes character when  the sign of $\omega(\muu_\infty,\lambda_\infty)$ changes. Thus a bifurcation occurs when $\omega(\muu_\infty,\lambda_\infty)=0$, or equivalently,
\begin{align}\label{eq:bif}
\Omega(\muu_\infty,\lambda_\infty)=0.
\end{align}
Note that $Q(\muu_\infty,\lambda_\infty)$ and $F(\muu_\infty,\lambda_\infty)$ represent respectively the functional dependence of $P$ and $N$ on $\muu_\infty$ and $\lambda_\infty$. Thus the above bifurcation condition is simply the vanishing of the Jacobian determinant of $P$ and $N$ as functions of $\muu_\infty$ and $\lambda_\infty$. This is consistent with the previous work \cite{fu2016localized} and \cite{yu2022analytic}.

We consider two typical loading scenarios: either the resultant axial force $N$  or the axial stretch at infinity $\lambda_\infty$ is fixed. The latter case is used to approximate the case of fixed axial length, which can be realized more easily experimentally or in Abaqus simulations.

Let us first assume that the resultant axial force  $N=N_\text{c}$ is fixed, where $N_\text{c}$ is the prescribed axial force. Denote by $(a_\text{cr},\lambda_\text{cr})$ the root of the system of equations
\begin{align}\label{eq:bifNc}
\omega(\muu_\infty,\lambda_\infty)=0,\quad F(\muu_\infty,\lambda_\infty)=N_\text{c},
\end{align}
at which the bifurcation occurs according to the previous discussion.
In the vicinity of the bifurcation point, the amplitude equation \eqref{eq:amp} reduces to
\begin{align}\label{eq:yZ}
D(a_\text{cr})y''(Z)=\omega'(a_\text{cr},\lambda_\text{cr})(\muu_\infty-a_\text{cr}) y(Z)+\gamma(a_\text{cr},\lambda_\text{cr})y(Z)^2,
\end{align}
where the prime on $\omega$ denotes $d/d\muu_\infty=\partial/\partial\muu_\infty+(\partial/\partial\lambda_\infty)(d\lambda_\infty/d\muu_\infty)$. The above equation admits a localized solution of the form
\begin{align}
y(Z)=-\frac{3\omega'(\muu_\text{cr},\lambda_\text{cr})}{2\gamma(\muu_\text{cr},\lambda_\text{cr})}(\muu_\infty-\muu_\text{cr})\sech^2\Big(\frac{1}{2}\sqrt{\frac{\omega'(a_\text{cr},\lambda_\text{cr})}{D(a_\text{cr})}(\muu_\infty-\muu_\text{cr})}Z\Big).
\end{align}

On the other hand, the weakly nonlinear amplitude equation derived from the 3d theory \citep{ye2020weakly} takes the form
\begin{align}\label{eq:cZ}
 c''_1(Z)=\lambda_\text{cr}^2k_1(\muu_\infty-\muu_\text{cr}) c_1(Z)+\lambda_\text{cr}^2 k_2c_1(Z)^2,
\end{align}
where $c_1(Z)$ and $y(Z)$ are related by
\begin{align}
y(Z)=k c_1(Z)
\end{align}
with $k=-{2\lambda(a) }/{\lambda'(a)}|_{a=a_{\text{cr}}}$,
and $k_1$ and $k_2$ are constants available in \cite{ye2020weakly}. One can see that \eqref{eq:yZ} and \eqref{eq:cZ} are identical provided
\begin{align} \la{k1k2}
k_1=\frac{\omega'(a_\text{cr},\lambda_\text{cr})}{\lambda_\text{cr}^2 D(a_\text{cr})},\quad k_2=\frac{ k \gamma(a_\text{cr},\lambda_\text{cr}) }{\lambda_\text{cr}^2D(a_\text{cr})}.
\end{align}
We have verified numerically that this is indeed the case, but the current expressions on the right-hand sides of \rr{k1k2} are more compact and revealing.

The case of fixed $\lambda_\infty$ can be handled similarly. Let $(\muu_\text{cr}, \lambda_\text{cr})$  be the solution to the system of equations
\begin{align}\label{eq:biflambda}
\omega(\muu_\infty,\lambda_\infty)=0,\quad \lambda_\infty=\lambda_\text{c},
\end{align}
where $\lambda_\text{c}$ is a given constant. In the vicinity of the bifurcation point,  the amplitude equation parallel to \eqref{eq:yZ} is of the form
\begin{align}
D(\muu_\text{cr})y''(Z)=\omega'(a_\text{cr},\lambda_\text{cr})(\muu_\infty-a_\text{cr}) y(Z)+\gamma(a_\text{cr},\lambda_\text{cr})y(Z)^2,
\end{align}
where  the prime on $\omega$ now signifies ${\partial}/{\partial\muu_\infty}$. Similar to the previous case, it can be verified that the above amplitude equation is the same as its counterparts in \cite{ye2020weakly}.

\section{Comparison with Abaqus simulations}\label{sec:numerical}
In this section, we demonstrate the power of the 1d model by applying it to investigate localized bulging in an inflated tube of finite wall thickness in the fully nonlinear regime. Previous studies on this problem usually treat the tube as a finite length tube, but the problem can be analyzed more easily and very accurately by assuming  the tube to be of infinite length. This assumption only fails when the tube is very short and when bulging is no longer localized in the axial direction \citep{wang2021necking}. The reason is that bulging solutions decay exponentially towards the two ends.  Thus in the following analysis, we shall assume that the tube is effectively infinite and focus on solutions  subject to decaying boundary conditions. This assumption is validated by comparison with Abaqus simulations based on tubes of finite lengths. We shall consider the two loading scenarios  discussed in Subsection \ref{subsec:weakly} and compare the predictions of the 1d model with Abaqus simulations, which allows us to quantify the accuracy of our 1d model and determine its range of validity.
In all numerical calculations and Abaqus simulations, we use the Gent material model
\begin{align}
W=-\frac{\mu}{2}J_m\ln\Big(1-\frac{\lambda_1^2+\lambda_2^2+\lambda_3^2-3}{J_m}\Big),
\end{align}
where $\mu$ is the shear modulus and $J_m$ is a material constant. The Gent material model is chosen because it is commonly adopted to model the latex rubber tubes used in inflation experiments \citep{wang2019experimental}. We  take $\mu=1$ which is equivalent to scaling all stress variables by $\mu$ and $J_m=97.2$ which is typical for rubber.  The geometry of the tube  is taken to be  $H/R_m=0.4$ and $2L/R_m=40$, where $R_m=(A+B)/2$ is the average radius. In the Abaqus simulations, to ensure that localized bulging occurs in the middle of the tube, a small section with length $0.1L$ around the middle point of the tube is weakened by taking its shear modulus to be $0.9999$ times that of the rest of the tube.

The 1d differential equation \eqref{eq:1d} subject to appropriate end conditions (see \eqref{eq:ini} later) can be solved numerically with the aid of the symbolic computation software {\it Mathematica}. Although the gradient modulus $D(\muu)$ involves an integral that cannot be evaluated analytically, this integral can be defined numerically in Mathematica with the built-in command \texttt{?NumericQ}  and can be manipulated as elementary functions. Numerically solving the 1d equation is significantly faster than Abaqus simulations. The 1d equation can typically be solved in a	 few seconds on a personal computer for the case of fixed axial force.

\subsection{The case of fixed axial force}

We first consider the loading scenario whereby the resultant axial force $N$ is fixed.  As mentioned earlier, we assume that the tube is infinitely long and focus on the solution that satisfies the decaying boundary condition
\begin{align}\label{eq:bc}
\lim_{Z\to \infty}\muu(Z)=\muu_\infty.
\end{align}
A linear analysis shows that the solution to \eqref{eq:el} satisfying \eqref{eq:bc} decays exponentially as $Z\to\infty$. Thus we  have $\lim_{Z\to \infty}\muu'(Z)=0$ automatically.
We assume that the bulging solution is symmetric with respect to $Z=0$ so that $\muu'(0)=0$. We write $\lambda_\infty=\lambda(\muu_\infty)$, $\muu_0=\muu(0)$ and $\lambda_0=\lambda(\muu(0))$. Since $(\muu_\infty,\lambda_\infty)$ satisfy  \rr{eq:eqQ} and \rr{eq:M}, we have
\begin{align}
&M(\muu_\infty,\lambda_\infty)-\frac{1}{2}\muu_\infty^2 Q(\muu_\infty,\lambda_\infty)-\frac{N}{2\pi A^2}=0, \label{eq:lambdainf}\\
&Q(\muu_\infty,\lambda_\infty)-P=0. \label{eq:PN}
\end{align}
From the definition of $\lambda_0$ and the conservation law \eqref{eq:first}, we see that
 $(\muu_0,\lambda_0)$ satisfies
\begin{align}
&M(\muu_0,\lambda_0)-\frac{1}{2}\muu_0^2Q(\muu_\infty,\lambda_\infty)=M(\muu_\infty,\lambda_\infty)-\frac{1}{2}\muu_\infty^2 Q(\muu_\infty,\lambda_\infty),\label{eq:mu00}\\
&G(\muu_0,\lambda_0)=G(\muu_\infty,\lambda_\infty).\label{eq:mu0}
\end{align}
Either $\muu_\infty$ or $P$ can be taken to be the load parameter.
When $\muu_\infty$ is specified, one can first obtain $\lambda_\infty$ from \eqref{eq:lambdainf}. The associated $P$ is computed according to \rr{eq:PN}. Then solving \eqref{eq:mu00} and \eqref{eq:mu0} for nontrivial solutions, one obtains the \lq\lq initial" values $\muu_0$ and  $\lambda_0$. The localized solution can be found  by solving  the initial value problem
\begin{align}
&A^2\muu\lambda(\muu)(Q(\muu,\lambda(\muu))-P)-\frac{1}{2}D'(\muu)\muu'(Z)^2-D(\muu)\muu''(Z)=0, \label{eq:ini}\\
&\muu(0)=\muu_0,\quad \muu'(0)=0. \label{eq:inibc}
\end{align}
As a first example,  fixing the axial force $N$ to be zero,  we find from the bifurcation condition \eqref{eq:bifNc} that localized bulging takes place at $\muu_\infty=\muu_\text{cr}=1.86$ with a critical pressure $P_\text{cr}=0.308$. As we trace the bifurcation solution away from the bifurcation point, the pressure drops while the bulge grows until it has almost reached a maximum amplitude after which the bulge will propagate at a constant pressure.  From Maxwell's equal-areal rule, the propagation pressure  is $P_M=0.197$.

\begin{figure}[htbp!]
	\centering
	\subfloat[]{\includegraphics[width=0.42\textwidth]{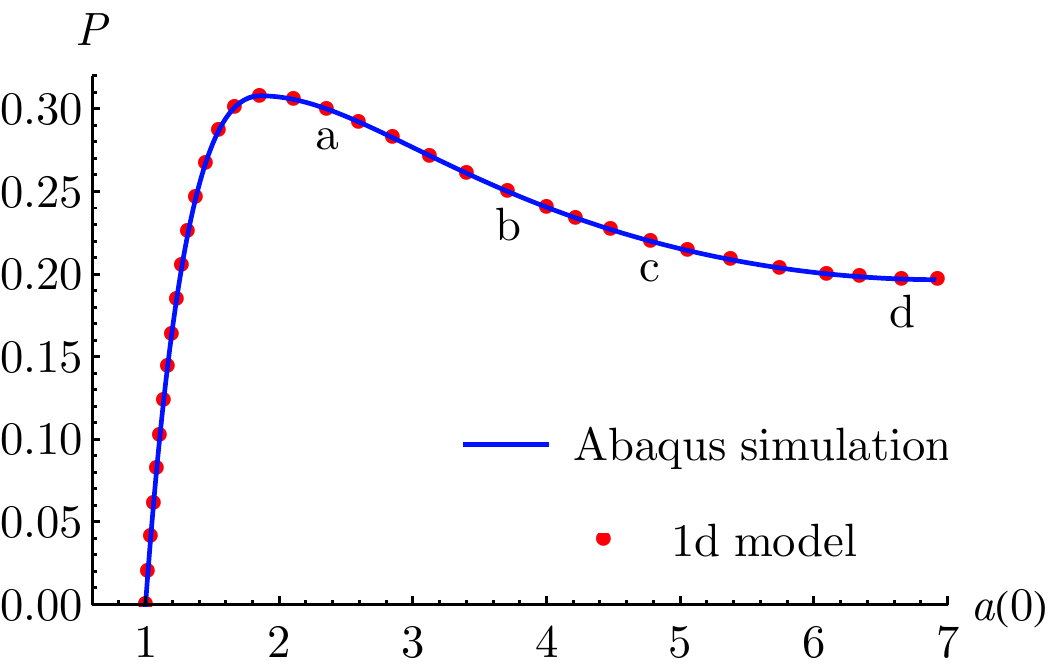}
	}\qquad
	\subfloat[]{\includegraphics[width=0.42\textwidth]{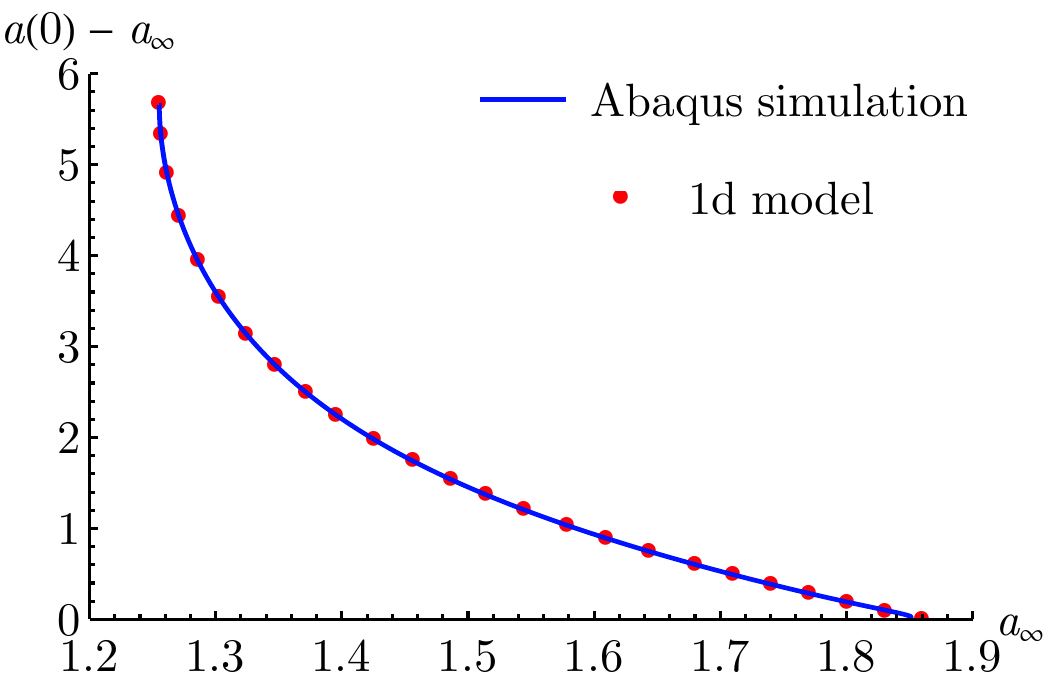}}
	\caption{Dependence of (a) pressure on  $\muu(0)$ and (b) bulging amplitude on  $\muu_\infty$, based on Abaqus simulations and the 1d model \rr{eq:ini} and \rr{eq:inibc} for fixed $N=0$. (Online version in color.)}
	\label{fig:N}
\end{figure}

\begin{figure}[htbp!]
	\centering
	\subfloat[]{\includegraphics[width=0.41\textwidth]{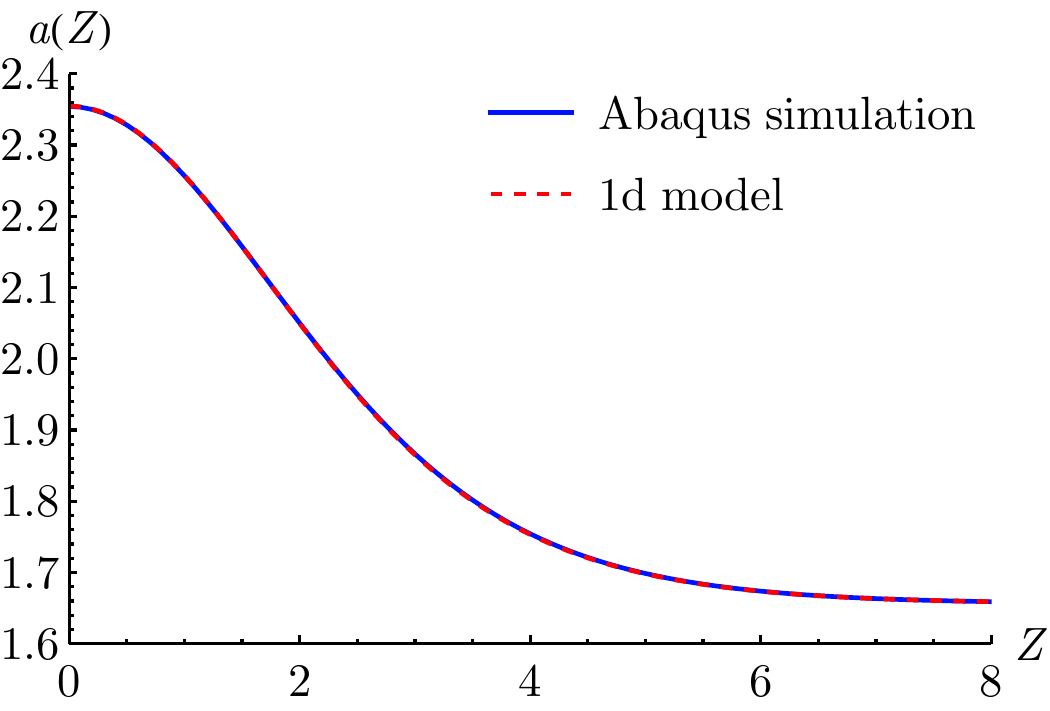}
	}\qquad
	\subfloat[]{\includegraphics[width=0.41\textwidth]{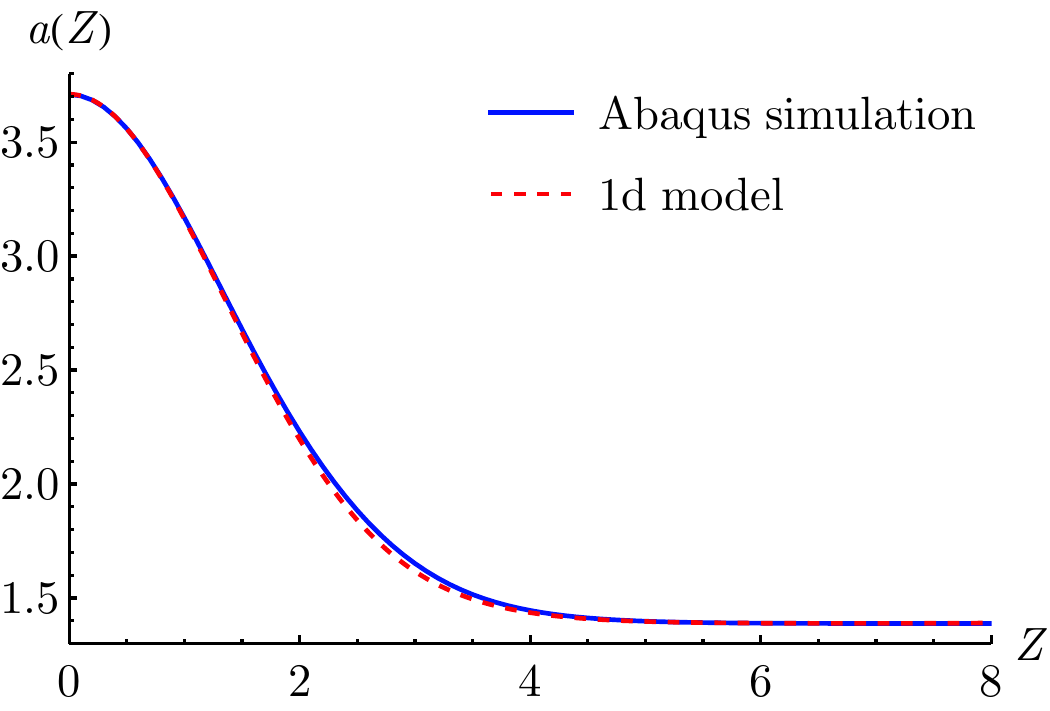}
	}\\
  \ \subfloat[]{\includegraphics[width=0.41\textwidth]{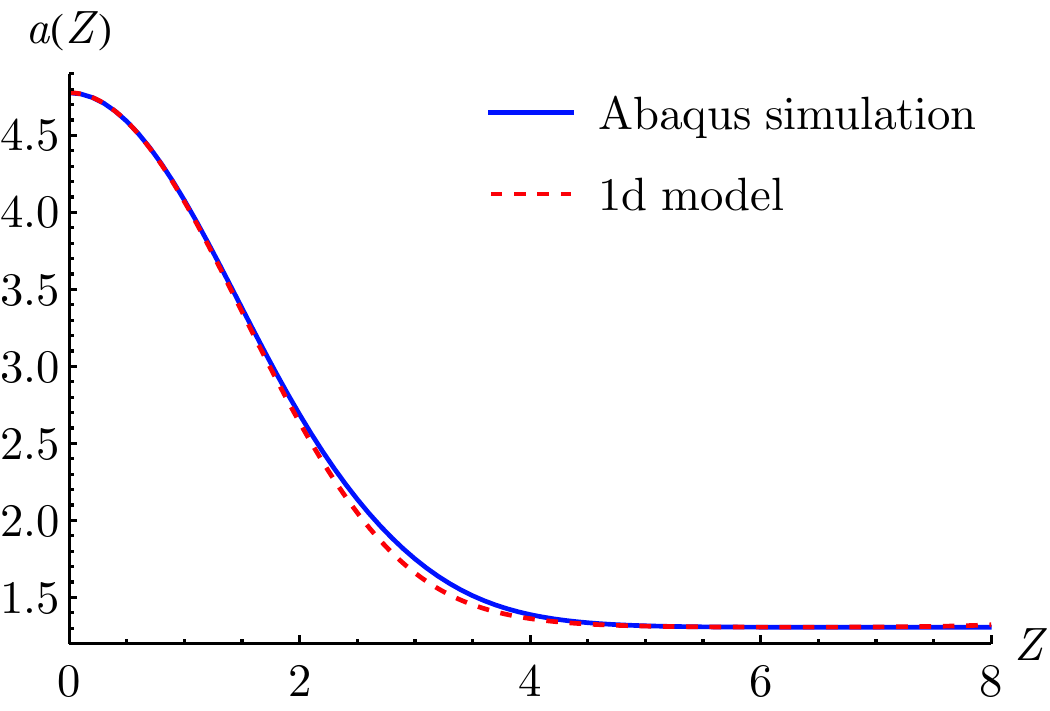}
    }\qquad \
    \subfloat[]{\includegraphics[width=0.41\textwidth]{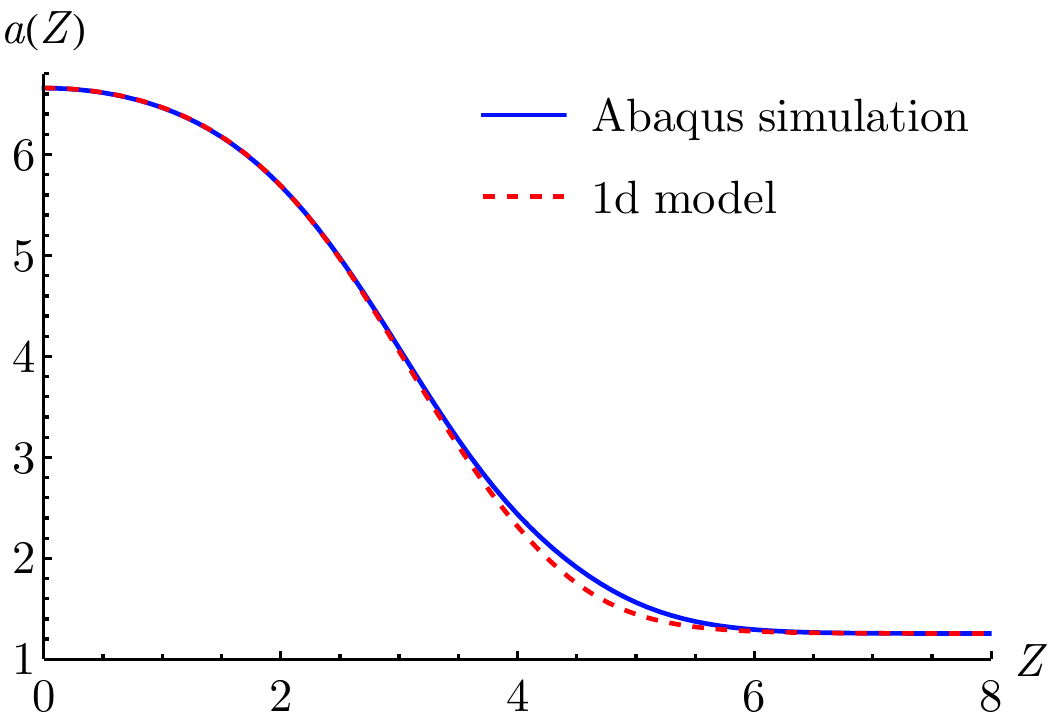}
    }
	\caption{Bulging solutions given by Abaqus simulations and the 1d model  at the four states marked in Fig. \ref{fig:N}(a)  for fixed  $N=0$: (a) $P=0.3$, (b) $P=0.25$, (c) $P=0.22$, (d) $P=0.197$. (Online version in color.)}
	\label{fig:Nc}
\end{figure}

Fig.~\ref{fig:N} shows the dependence of the pressure on  $\muu(0)$ and the bulging amplitude on  $\muu_\infty$ based on Abaqus simulations and use of the 1d model. The bulging solutions given by Abaqus simulations and the 1d model at the four states marked in Fig. \ref{fig:N}(a)  are shown in Fig. \ref{fig:Nc}. It is seen that the 1d solution agrees well with Abaqus simulations in the entire post-bifurcation regime. Remarkably, the 1d solution remains highly accurate even in the final propagation stage, as shown in Fig. \ref{fig:Nc}(d). Note also that the Abaqus simulations and 1d calculations are conducted for $2 L=40 R_m$ and $\infty$, respectively. This verifies our earlier claim that the tube can effectively be viewed to be infinitely long.

\subsection{The case of fixed ends}
Next, we consider the loading scenario whereby the tube is first stretched to a specified length $2\ell$ and then its two ends are fixed to prevent further axial displacement (whether the radial displacement is restricted or not at the ends is immaterial since the tube is assumed to be sufficiently long). In the previous subsection, we have solved the problem for a specified axial force $N$ or equivalently a specified $\lambda_\infty$. For the current problem with a given $\ell$, we define $\lambda_\text{c}=\ell/L$ and we need to find $\lambda_\infty$ such that  the following condition is satisfied:
\begin{align}\label{eq:target}
\int_0^L \lambda(\muu(Z))\,dZ=\lambda_\text{c} L.
\end{align}
This can be achieved by a shooting procedure: for each $\lambda_\infty$, we compute the left-hand side using the procedure outlined in the previous subsection and adjust $\lambda_\infty$ such that the left-hand side and the right-hand side are equal. The procedure may be started by taking $\lambda_\infty={\lambda}_\text{c}$.  However, solving the present problem  by the shooting procedure requires a lot of adjustments by hand due to the fact that the localized solutions that we are looking for are extremely close to periodic solutions. To find solutions for the current case in a more robust way, we use the finite difference method instead.

To implement the finite difference method, we partition the domain $[0,L]$ using a uniform mesh $Z_0$, $Z_1$, $\dots$, $Z_n$ with mesh size $h=L/n$ and coordinate of the $j$-th grid point given by $Z_j=jh$. We use $\muu_j$ to represent the numerical approximation of $\muu(Z_j)$. Applying the central difference scheme, we convert the differential equation \eqref{eq:ini} into a set of algebraic equations
\begin{align}
\begin{split}
&A^2\muu_j \lambda(\muu_j)(Q(\muu_j,\lambda(\muu_j))-P)-\frac{1}{2}D'(\muu_j)\Big(\frac{\muu_{j+1}-\muu_{j-1}}{2h}\Big)^2\\
&-D(\muu_j)\frac{\muu_{j+1}-2\muu_j+\muu_{j-1}}{h^2}=0,\quad j=1,2,\dots,n-1. \label{eq:diff}
\end{split}
\end{align}
The left boundary condition is given by
\begin{align}
\muu'(0)=0.
\end{align}
We see from \eqref{eq:amp} that the solution to \eqref{eq:ini} subject to \eqref{eq:bc} has the asymptotic behavior
\begin{align}
a(Z) \sim a_\infty + a_1 e^{-\kappa Z}\quad \text{as}\ Z\to \infty,
\end{align}
where $a_1$ is a constant and
$$ \kappa=\sqrt{\frac{\omega(a_\infty,\lambda_\infty)}{D(a_\infty)}}. $$
Because of this, we may replace the decaying boundary condition  \eqref{eq:bc} by the ``soft" asymptotic condition
\begin{align}
	\muu'(L)+\kappa(\muu(L)-\muu_\infty)=0.
\end{align}
To avoid the loss of accuracy at the two endpoints, we introduce two additional unknowns $\muu_{-1}$ and $\muu_{n+1}$. Then the left and right boundary conditions yield
\begin{align}
&\frac{\muu_1-\muu_{-1}}{2h}=0, \label{eq:mu-1}\\
&\frac{\muu_{n+1}-\muu_{n-1}}{2h}+\kappa (\muu_n-\muu_\infty)=0. \label{eq:mun+1}
\end{align}
Solving for $\muu_{-1}$ and $\muu_{n+1}$ from the above equations, and substituting them into the finite difference equation \eqref{eq:diff} evaluated at $j=0$ and $j=n$, we obtain the following discrete boundary conditions with truncation errors of order $h^2$:
\begin{align}
&A^2\muu_0 \lambda(\muu_0)(Q(\muu_0,\lambda(\muu_0))-P)-2D(\muu_0)\frac{\muu_{1}-\muu_0}{h^2}=0, \label{eq:diffbc1}\\
\begin{split}
&A^2\muu_n \lambda(\muu_n)(Q(\muu_n,\lambda(\muu_n))-P)-\frac{1}{2}D'(\muu_n)\kappa^2 (\muu_n-\muu_\infty)^2\\
&-2D(\muu_n)\frac{\muu_{n-1}-\muu_n-h\kappa (\muu_n-\muu_\infty)}{h^2}=0. \label{eq:diffbc2}
\end{split}
\end{align}
Finally, the fixed-length restriction \eqref{eq:target} gives
\begin{align}\label{eq:ell}
\frac{1}{2}\lambda(\muu_0)+\sum_{j=1}^{n-1}\lambda(\muu_j)+\frac{1}{2}\lambda(\muu_n)-\frac{\lambda_\text{c}L}{h} =0.
\end{align}

For the current loading scenario,  one can still use $a_\infty$ or $P$ as the loading parameter. However, it is more convenient to choose $a_0$ as the loading parameter since it is monotonically increasing during inflation, and treat $a_\infty$ and $\lambda_\infty$ as unknowns. We see from \eqref{eq:lambdainf} that $N$ is a function of $a_\infty$ and $\lambda_\infty$. It follows that $\lambda(\mu)$ and $D(\mu)$ also depend on $a_\infty$ and $\lambda_\infty$ through their dependence on $N$. This implicit dependence should be considered when solving the above algebraic equations.

Given $a_0$, setting $n$ to be a sufficiently large number and solving the system of $(n+2)$ algebraic equations consisting of \eqref{eq:diff}, \eqref{eq:diffbc1}, \eqref{eq:diffbc2} and \eqref{eq:ell} for $\muu_j$, $1\leq j\leq n$, $a_\infty$ and $\lambda_\infty$ with a suitable initial guess, we obtain the finite-difference solution for the present problem. We may use the weakly nonlinear solution with $\lambda_\infty=\lambda_\text{c}=\ell/L$ as an initial guess in the near-critical regime and continue the solution to the fully nonlinear regime by always using the solution at the previous step as the initial guess for the current step.

\begin{figure}[h!]
	\centering
	\subfloat[]{\includegraphics[width=0.42\textwidth]{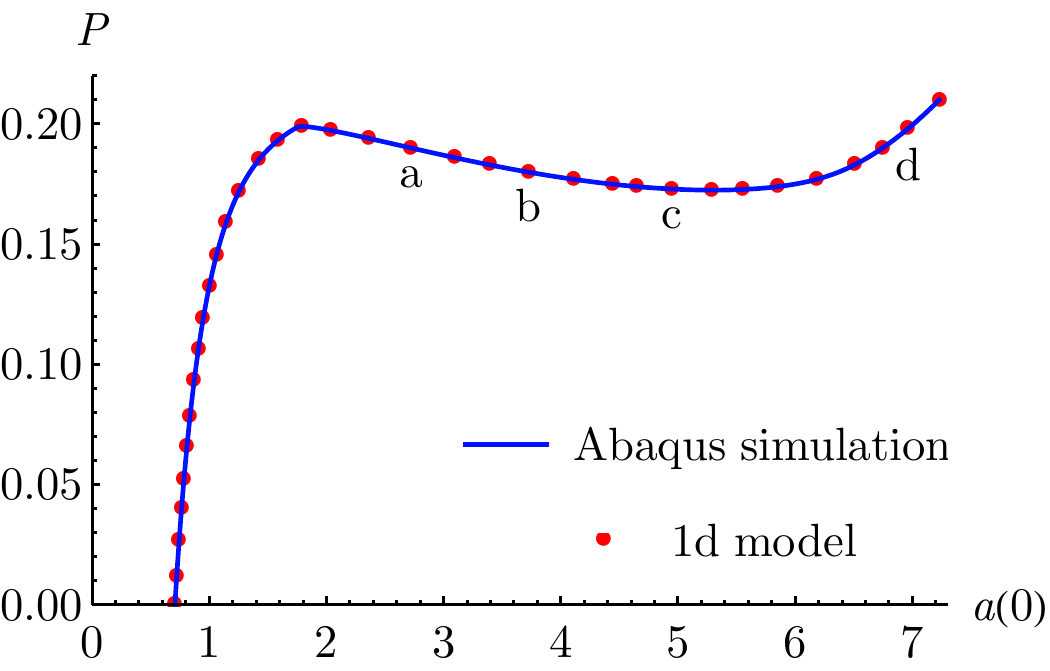}
	}\qquad
	\subfloat[]{\includegraphics[width=0.42\textwidth]{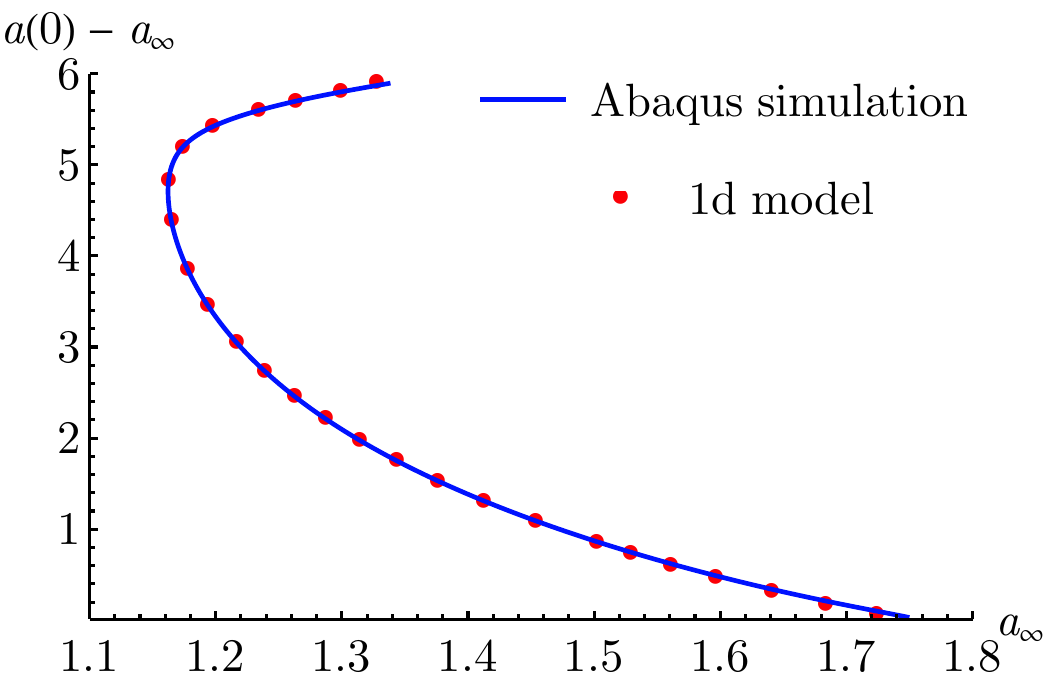}}
	\caption{ Dependence of (a)  pressure on $\muu(0)$ and (b) bulging amplitude on $\muu_\infty$, based on Abaqus simulations and the 1d model with finite difference scheme for fixed length $\ell/L=2$. (Online version in color.) }
	\label{fig:l}
\end{figure}

\begin{figure}[h!]
	\centering
	\subfloat[]{\includegraphics[width=0.41\textwidth]{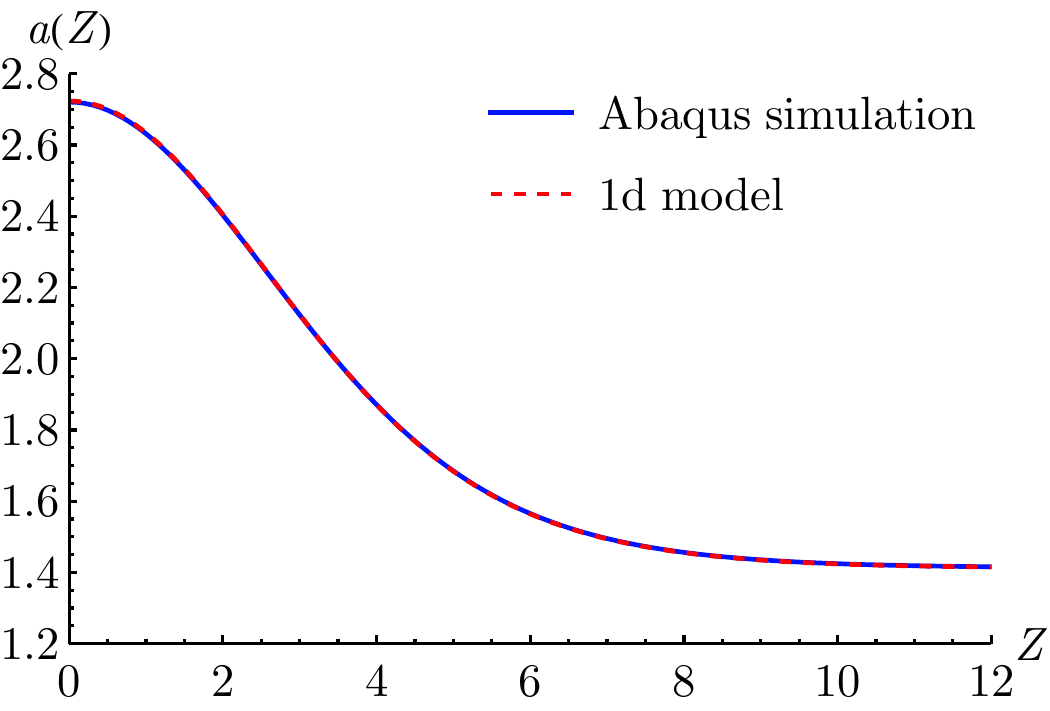}	
	}\qquad
	\subfloat[]{\includegraphics[width=0.41\textwidth]{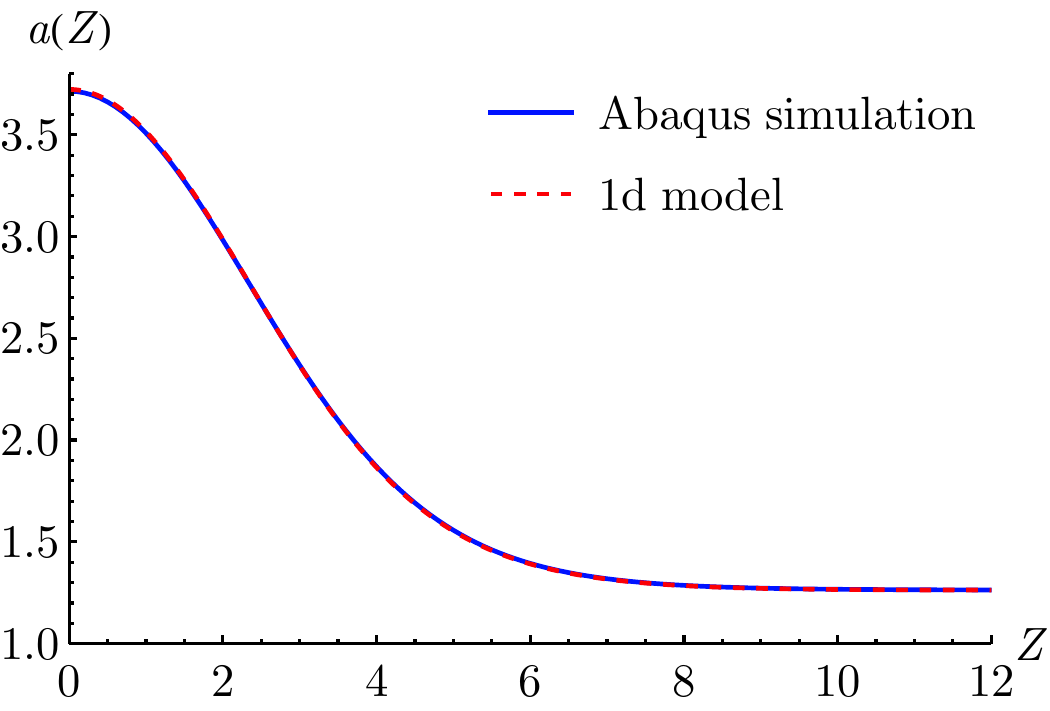}
	}\\
	\quad \subfloat[]{\includegraphics[width=0.41\textwidth]{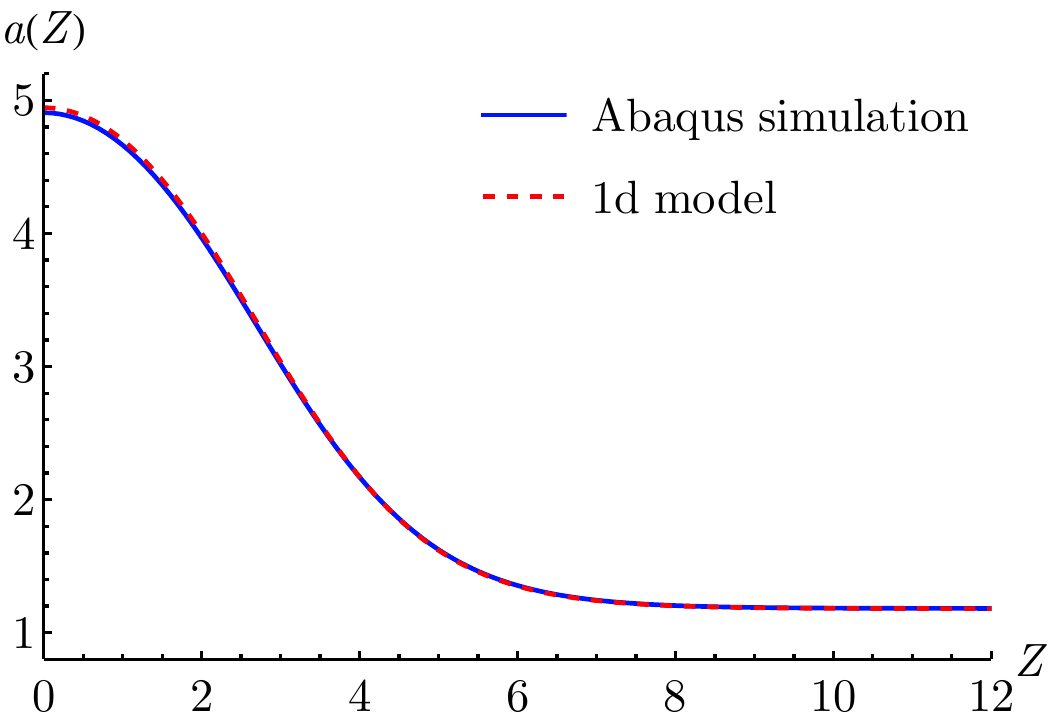}
	}\qquad
	\subfloat[]{\includegraphics[width=0.41\textwidth]{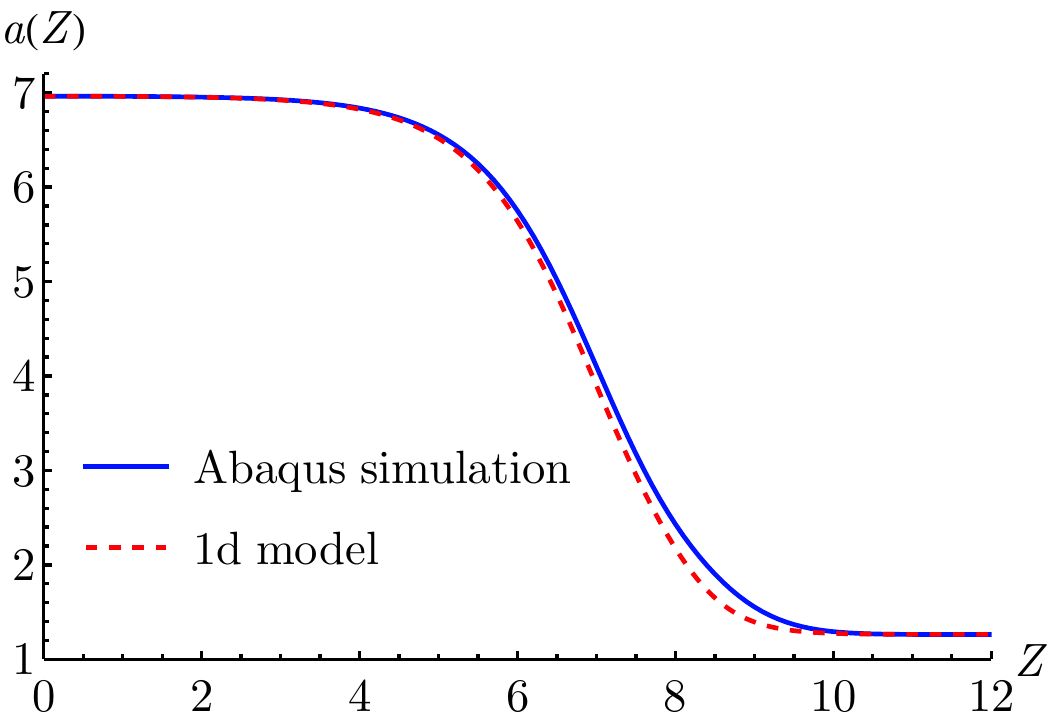}
	}
	\caption{Bulging solutions based on Abaqus simulations and the 1d model at the four states indicated in Fig. \ref{fig:l}(a) for fixed length $\ell/L=2$: (a) $P=0.19$,  (b) $P=0.18$, (c) $P=0.173$, (d) $P=0.198$. (Online version in color.)}
	\label{fig:ll}
\end{figure}

When the total length is fixed to be $\ell=2L$, then initially $\lambda_\infty=2$  and localized bulging takes place at $\muu_\infty=\muu_\text{cr}=1.74$ with a critical pressure $P_\text{cr}=0.198$ according to \eqref{eq:biflambda}. In Fig. \ref{fig:l}, we have shown the dependence of the pressure on $\muu(0)$ and the bulging amplitude on $\muu_\infty$ based on  Abaqus simulations and use of the 1d model. The  bulging solutions determined by Abaqus simulations and the 1d model at the four states indicated in Fig. \ref{fig:l}(a) are presented in Fig. \ref{fig:ll}. It is observed that the agreement between the 1d model and  Abaqus simulations is again excellent in the fully nonlinear regime.

\begin{figure}[h!]
	\centering
	\subfloat[]{\includegraphics[width=0.42\textwidth]{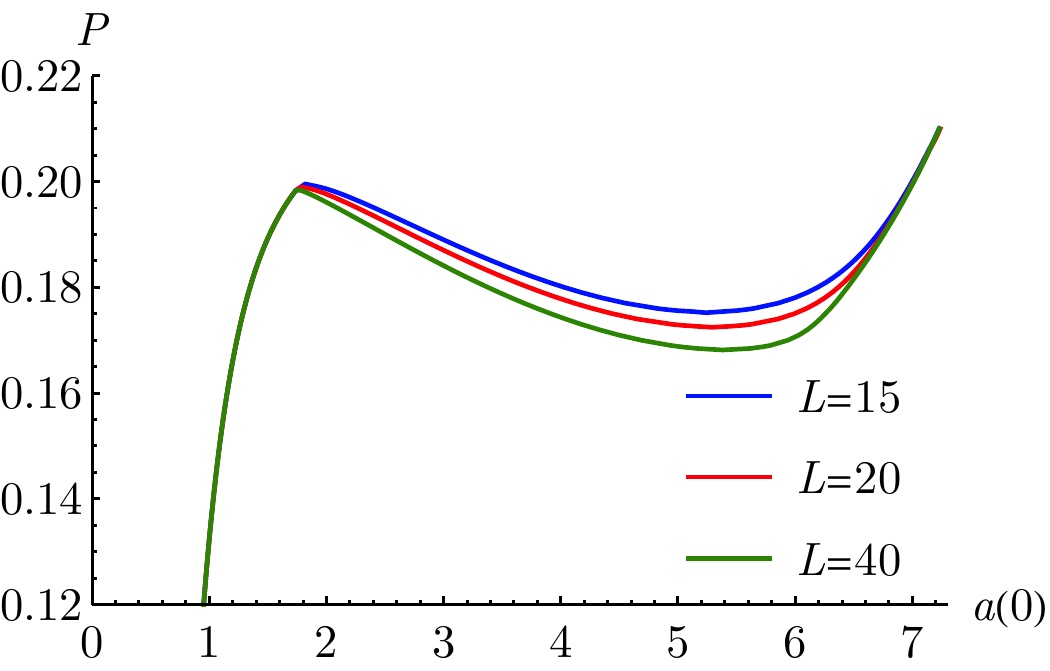}
	}\qquad
	\subfloat[]{\includegraphics[width=0.42\textwidth]{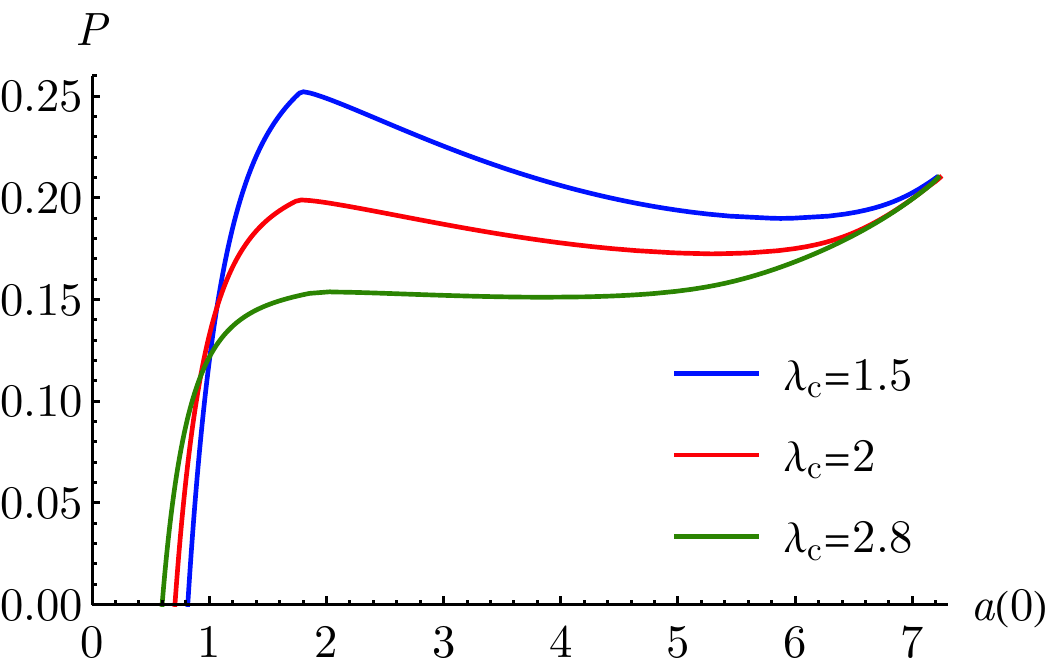}
	}
	\caption{Variation of $P$ against $\muu(0)$ predicted by the 1d model when the total length is fixed during inflation: (a) $\lambda_\text{c}=2$ and $L=15$, $20$ and $40$, respectively, and (b) $L=20$ and $\lambda_\text{c}=1.5$, $2$ and $2.8$, respectively. (Online version in color.)}
	\label{P-mu0-varying}
\end{figure}

Finally,  Fig.~\ref{P-mu0-varying} shows the actual variation of $P$ against $a(0)$ predicted by the 1d model when (a) the averaged stretch $\lambda_\text{c}$ is fixed and $L$ is varied, or (b) $L$ is fixed but $\lambda_\text{c}$ is varied. These results confirm the theoretical prediction of \cite{guo2022localized} that the right branches of these curves all converge to a master curve that is independent of $L$ or $\lambda_\text{c}$. These curves all terminate at the point where the axial stress near each end of the tube has become compressive enough so that secondary  Euler buckling or axisymmetric wrinkling becomes possible.

\section{Conclusion}\label{sec:con}

We have derived a 1d model for the analysis of axisymmetric deformations of an inflated cylindrical tube of finite wall thickness, and established its range of validity by comparing its predictions with those of Abaqus simulations for two typical loading scenarios. The comparison shows that the 1d model performs extremely well in both the near-critical and fully nonlinear regimes. The dimension reduction started from three-dimensional finite elasticity theory and is performed in terms of the energy functional and principal stretches. A key ingredient of the dimension reduction is the assumption of slow variation of the leading-order solution in the axial direction without any restriction on its amplitude, which results in a 1d model that is simple but is still capable of capturing the strain-gradient effect. This is in contrast with the traditional asymptotic analysis where the leading order solution is assumed to be a small-amplitude perturbation from the primary deformation. It is because of this difference that  the 1d model has a much larger range of validity than the expansion methods around the bifurcation point. The nonlinearity of the strain  is kept in the 1d model, reflected by the nonlinear potential $G(\muu,\lambda(\muu))$ and the nonlinear strain-gradient modulus $D(\muu)$. Our expression for the strain-gradient coefficient $D(\muu)$ is quite simple. For the Gent material model, $D(\muu)$ can be calculated by integrating once. We remark that although the derivation presented in this work is variational, the 1d model can also be derived by substituting the asymptotic solution \eqref{eq:zrp} into the 3d governing equations and solving the resulting equations at successive orders.

The 1d model is amenable to asymptotic and numerical solutions. The bifurcation condition and the weakly nonlinear amplitude equation predicted by the 1d model are exact. In fact, the expressions \rr{eq:omega} and \rr{eq:omega1} derived using the 1d model are more compact and more revealing than their counterparts in \cite{ye2020weakly}. A major advantage of the 1d model is that the entire evolution process of bulging or necking can be determined using the finite difference method which is more accessible and much easier to implement than commercial packages such as Abaqus. This advantage would become even more significant when other fields such as electric  loadings and residual stresses were also present. Such extra fields and new geometries (e.g. electric field \citep{fu2018localized}, axisymmetric necking of a stretched plate \citep{wjf2022} or their combination \citep{fu2023axisymmetric}) will be considered in our future studies.

A Mathematica code that produces all the results presented in the paper is available on GitHub (\url{https://github.com/yfukeele}).

\section*{Acknowledgments}
This work was supported by the National Natural Science Foundation of China (Grant No 12072224) and the Engineering and Physical Sciences Research Council, UK (Grant No EP/W007150/1).


\begin{thebibliography}{56}
	\expandafter\ifx\csname natexlab\endcsname\relax\def\natexlab#1{#1}\fi
	\providecommand{\url}[1]{\texttt{#1}}
	\providecommand{\href}[2]{#2}
	\providecommand{\path}[1]{#1}
	\providecommand{\DOIprefix}{doi:}
	\providecommand{\ArXivprefix}{arXiv:}
	\providecommand{\URLprefix}{URL: }
	\providecommand{\Pubmedprefix}{pmid:}
	\providecommand{\doi}[1]{\href{http://dx.doi.org/#1}{\path{#1}}}
	\providecommand{\Pubmed}[1]{\href{pmid:#1}{\path{#1}}}
	\providecommand{\bibinfo}[2]{#2}
	\ifx\xfnm\relax \def\xfnm[#1]{\unskip,\space#1}\fi
	\bibitem[{Alhayani et~al.(2014)Alhayani, Rodr\'iguez \& Merodio}]{arm2014}
	\bibinfo{author}{Alhayani, A.~A.}, \bibinfo{author}{Rodr\'iguez, J.}, \&
	\bibinfo{author}{Merodio, J.} (\bibinfo{year}{2014}).
	\newblock \bibinfo{title}{Competition between radial expansion and axial
		propagation in bulging of inflated cylinders with application to aneurysms
		propagation in arterial wall tissue}.
	\newblock {\it \bibinfo{journal}{Int. J. Eng. Sci.}\/},  {\it
		\bibinfo{volume}{85}\/}, \bibinfo{pages}{74--89}.
	\bibitem[{Althobaiti(2022)}]{althobaiti2022effect}
	\bibinfo{author}{Althobaiti, A.} (\bibinfo{year}{2022}).
	\newblock \bibinfo{title}{Effect of torsion on the initiation of localized
		bulging in a hyperelastic tube of arbitrary thickness}.
	\newblock {\it \bibinfo{journal}{Z. fur Angew. Math. Phys.}\/},  {\it
		\bibinfo{volume}{73}\/}, \bibinfo{pages}{1--11}.
	\bibitem[{Audoly \& Hutchinson(2016)}]{audoly2016analysis}
	\bibinfo{author}{Audoly, B.}, \& \bibinfo{author}{Hutchinson, J.~W.}
	(\bibinfo{year}{2016}).
	\newblock \bibinfo{title}{Analysis of necking based on a one-dimensional
		model}.
	\newblock {\it \bibinfo{journal}{J. Mech. Phys. Solids}\/},  {\it
		\bibinfo{volume}{97}\/}, \bibinfo{pages}{68--91}.
	\bibitem[{Audoly \& Neukirch(2021)}]{audoly2021one}
	\bibinfo{author}{Audoly, B.}, \& \bibinfo{author}{Neukirch, S.}
	(\bibinfo{year}{2021}).
	\newblock \bibinfo{title}{A one-dimensional model for elastic ribbons: a little
		stretching makes a big difference}.
	\newblock {\it \bibinfo{journal}{J. Mech. Phys. Solids}\/},  {\it
		\bibinfo{volume}{153}\/}, \bibinfo{pages}{104457}.
	\bibitem[{Bastola \& Hossain(2021)}]{bastola2021shape}
	\bibinfo{author}{Bastola, A.~K.}, \& \bibinfo{author}{Hossain, M.}
	(\bibinfo{year}{2021}).
	\newblock \bibinfo{title}{The shape--morphing performance of magnetoactive soft
		materials}.
	\newblock {\it \bibinfo{journal}{Mater. Des.}\/},  {\it
		\bibinfo{volume}{211}\/}, \bibinfo{pages}{110172}.
	\bibitem[{Bucchi \& Hearn(2013)}]{bh2013}
	\bibinfo{author}{Bucchi, A.}, \& \bibinfo{author}{Hearn, G.~E.}
	(\bibinfo{year}{2013}).
	\newblock \bibinfo{title}{Delay or removal of aneurysm formation in the
		anaconda wave energy extraction device}.
	\newblock {\it \bibinfo{journal}{Renewable Energy}\/},  {\it
		\bibinfo{volume}{55}\/}, \bibinfo{pages}{104--119}.
	\bibitem[{Chater \& Hutchinson(1984)}]{ch1984}
	\bibinfo{author}{Chater, E.}, \& \bibinfo{author}{Hutchinson, J.~W.}
	(\bibinfo{year}{1984}).
	\newblock \bibinfo{title}{On the propagation of bulges and buckles}.
	\newblock {\it \bibinfo{journal}{ASME J. Appl. Mech.}\/},  {\it
		\bibinfo{volume}{51}\/}, \bibinfo{pages}{269--277}.
	\bibitem[{Collins et~al.(2021)Collins, Hossain, Dettmer \&
		Masters}]{collins2021flexible}
	\bibinfo{author}{Collins, I.}, \bibinfo{author}{Hossain, M.},
	\bibinfo{author}{Dettmer, W.}, \& \bibinfo{author}{Masters, I.}
	(\bibinfo{year}{2021}).
	\newblock \bibinfo{title}{Flexible membrane structures for wave energy
		harvesting: A review of the developments, materials and computational
		modelling approaches}.
	\newblock {\it \bibinfo{journal}{Renew. Sust. Energ. Rev.}\/},  {\it
		\bibinfo{volume}{151}\/}, \bibinfo{pages}{111478}.
	\bibitem[{Demirkoparan \& Merodio(2017)}]{dm2015}
	\bibinfo{author}{Demirkoparan, H.}, \& \bibinfo{author}{Merodio, J.}
	(\bibinfo{year}{2017}).
	\newblock \bibinfo{title}{Bulging bifurcation of inflated circular cylinders of
		doubly fiber-reinforced hyperelastic material under axial loading and
		swelling}.
	\newblock {\it \bibinfo{journal}{Math. Mech. Solids}\/},  {\it
		\bibinfo{volume}{22}\/}, \bibinfo{pages}{666--682}.
	\bibitem[{Emery(2023)}]{emery2023}
	\bibinfo{author}{Emery, D.} (\bibinfo{year}{2023}).
	\newblock \bibinfo{title}{Elasto-capillary necking, bulging and maxwell states
		in soft compressible cylinders}.
	\newblock {\it \bibinfo{journal}{Int. J. Non-linear Mech.}\/},  {\it
		\bibinfo{volume}{148}\/}, \bibinfo{pages}{104276}.
	\bibitem[{Emery \& Fu(2021{\natexlab{a}})}]{ef2021ab}
	\bibinfo{author}{Emery, D.}, \& \bibinfo{author}{Fu, Y.~B.}
	(\bibinfo{year}{2021}{\natexlab{a}}).
	\newblock \bibinfo{title}{Elasto-capillary circumferential buckling of soft
		tubes under axial loading: existence and competition with localised beading
		and periodic axial modes}.
	\newblock {\it \bibinfo{journal}{Mech. Soft Mater.}\/},  {\it
		\bibinfo{volume}{3}\/}, \bibinfo{pages}{3}.
	\bibitem[{Emery \& Fu(2021{\natexlab{b}})}]{ef2021a}
	\bibinfo{author}{Emery, D.}, \& \bibinfo{author}{Fu, Y.~B.}
	(\bibinfo{year}{2021}{\natexlab{b}}).
	\newblock \bibinfo{title}{Localised bifurcation in soft cylindrical tubes under
		axial stretching and surface tension}.
	\newblock {\it \bibinfo{journal}{Int. J. Solids Struct.}\/},  {\it
		\bibinfo{volume}{219-220}\/}, \bibinfo{pages}{23--33}.
	\bibitem[{Emery \& Fu(2021{\natexlab{c}})}]{ef2021c}
	\bibinfo{author}{Emery, D.}, \& \bibinfo{author}{Fu, Y.~B.}
	(\bibinfo{year}{2021}{\natexlab{c}}).
	\newblock \bibinfo{title}{Post-bifurcation behaviour of elasto-capillary
		necking and bulging in soft tubes}.
	\newblock {\it \bibinfo{journal}{Proc. R. Soc. A}\/},  {\it
		\bibinfo{volume}{477}\/}, \bibinfo{pages}{20210311}.
	\bibitem[{Fu et~al.(2018)Fu, Dorfmann \& Xie}]{fu2018localized}
	\bibinfo{author}{Fu, Y.~B.}, \bibinfo{author}{Dorfmann, L.}, \&
	\bibinfo{author}{Xie, Y.} (\bibinfo{year}{2018}).
	\newblock \bibinfo{title}{Localized necking of a dielectric membrane}.
	\newblock {\it \bibinfo{journal}{Extreme Mech. Lett.}\/},  {\it
		\bibinfo{volume}{21}\/}, \bibinfo{pages}{44--48}.
	\bibitem[{Fu \& Il'ichev(2015)}]{fi2015}
	\bibinfo{author}{Fu, Y.~B.}, \& \bibinfo{author}{Il'ichev, A.~T.}
	(\bibinfo{year}{2015}).
	\newblock \bibinfo{title}{Localized standing waves in a hyperelastic membrane
		tube and their stabilization by a mean flow}.
	\newblock {\it \bibinfo{journal}{Maths Mech. Solids}\/},  {\it
		\bibinfo{volume}{20}\/}, \bibinfo{pages}{1198--1214}.
	\bibitem[{Fu et~al.(2021)Fu, Jin \& Goriely}]{fjg2021}
	\bibinfo{author}{Fu, Y.~B.}, \bibinfo{author}{Jin, L.~S.}, \&
	\bibinfo{author}{Goriely, A.} (\bibinfo{year}{2021}).
	\newblock \bibinfo{title}{Necking, beading, and bulging in soft elastic
		cylinders}.
	\newblock {\it \bibinfo{journal}{J. Mech. Phys. Solids}\/},  {\it
		\bibinfo{volume}{147}\/}, \bibinfo{pages}{104250}.
	\bibitem[{Fu et~al.(2016)Fu, Liu \& Francisco}]{fu2016localized}
	\bibinfo{author}{Fu, Y.~B.}, \bibinfo{author}{Liu, J.~L.}, \&
	\bibinfo{author}{Francisco, G.~S.} (\bibinfo{year}{2016}).
	\newblock \bibinfo{title}{Localized bulging in an inflated cylindrical tube of
		arbitrary thickness--the effect of bending stiffness}.
	\newblock {\it \bibinfo{journal}{J. Mech. Phys. Solids}\/},  {\it
		\bibinfo{volume}{90}\/}, \bibinfo{pages}{45--60}.
	\bibitem[{Fu et~al.(2008)Fu, Pearce \& Liu}]{fu2008post}
	\bibinfo{author}{Fu, Y.~B.}, \bibinfo{author}{Pearce, S.~P.}, \&
	\bibinfo{author}{Liu, K.-K.} (\bibinfo{year}{2008}).
	\newblock \bibinfo{title}{Post-bifurcation analysis of a thin-walled
		hyperelastic tube under inflation}.
	\newblock {\it \bibinfo{journal}{Int. J. Non-Linear Mech.}\/},  {\it
		\bibinfo{volume}{43}\/}, \bibinfo{pages}{697--706}.
	\bibitem[{Fu et~al.(2012)Fu, Rogerson \& Zhang}]{frz2012}
	\bibinfo{author}{Fu, Y.~B.}, \bibinfo{author}{Rogerson, G.~A.}, \&
	\bibinfo{author}{Zhang, Y.~T.} (\bibinfo{year}{2012}).
	\newblock \bibinfo{title}{Initiation of aneurysms as a mechanical bifurcation
		phenomenon}.
	\newblock {\it \bibinfo{journal}{Int. J. Non-linear Mech.}\/},  {\it
		\bibinfo{volume}{47}\/}, \bibinfo{pages}{179--184}.
	\bibitem[{Fu \& Xie(2010)}]{fu2010stability}
	\bibinfo{author}{Fu, Y.~B.}, \& \bibinfo{author}{Xie, Y.~X.}
	(\bibinfo{year}{2010}).
	\newblock \bibinfo{title}{Stability of localized bulging in inflated membrane
		tubes under volume control}.
	\newblock {\it \bibinfo{journal}{Int. J. Eng. Sci.}\/},  {\it
		\bibinfo{volume}{48}\/}, \bibinfo{pages}{1242--1252}.
	\bibitem[{Fu \& Yu(2023)}]{fu2023axisymmetric}
	\bibinfo{author}{Fu, Y.~B.}, \& \bibinfo{author}{Yu, X.}
	(\bibinfo{year}{2023}).
	\newblock \bibinfo{title}{Axisymmetric necking of a circular electrodes-coated
		dielectric membrane}.
	\newblock {\it \bibinfo{journal}{arXiv preprint arXiv:2301.01129}\/}, .
	\bibitem[{Goncalves et~al.(2008)Goncalves, Pamplona \& Lopes}]{gp2008}
	\bibinfo{author}{Goncalves, P.~B.}, \bibinfo{author}{Pamplona, D.~C.}, \&
	\bibinfo{author}{Lopes, S. R.~X.} (\bibinfo{year}{2008}).
	\newblock \bibinfo{title}{Finite deformations of an initially stressed
		cylindrical shell under internal pressure}.
	\newblock {\it \bibinfo{journal}{Int. J. Mech. Sci.}\/},  {\it
		\bibinfo{volume}{50}\/}, \bibinfo{pages}{92--103}.
	\bibitem[{Green \& Adkins(1960)}]{gabook1960}
	\bibinfo{author}{Green, A.~E.}, \& \bibinfo{author}{Adkins, J.~E.}
	(\bibinfo{year}{1960}).
	\newblock {\it \bibinfo{title}{Large Elastic Deformations and Non-linear
			Continuum Mechanics}\/}.
	\newblock \bibinfo{publisher}{Clarendon Press, Oxford}.
	\bibitem[{Guo et~al.(2022)Guo, Wang \& Fu}]{guo2022localized}
	\bibinfo{author}{Guo, Z.~M.}, \bibinfo{author}{Wang, S.~B.}, \&
	\bibinfo{author}{Fu, Y.~B.} (\bibinfo{year}{2022}).
	\newblock \bibinfo{title}{Localised bulging of an inflated rubber tube with
		fixed ends}.
	\newblock {\it \bibinfo{journal}{Proc. R. Soc. A}\/},  {\it
		\bibinfo{volume}{380}\/}, \bibinfo{pages}{20210318}.
	\bibitem[{Haughton \& Ogden(1979)}]{ho1979b}
	\bibinfo{author}{Haughton, D.~M.}, \& \bibinfo{author}{Ogden, R.~W.}
	(\bibinfo{year}{1979}).
	\newblock \bibinfo{title}{Bifurcation of inflated circular cylinders of elastic
		material under axial loading ii. exact theory for thick-walled tubes}.
	\newblock {\it \bibinfo{journal}{J. Mech. Phy. Solids}\/},  {\it
		\bibinfo{volume}{27}\/}, \bibinfo{pages}{489--512}.
	\bibitem[{Hejazi et~al.(2021)Hejazi, Hsiang \& Srikantha~Phani}]{hhp2021}
	\bibinfo{author}{Hejazi, M.}, \bibinfo{author}{Hsiang, Y.}, \&
	\bibinfo{author}{Srikantha~Phani, A.} (\bibinfo{year}{2021}).
	\newblock \bibinfo{title}{Fate of a bulge in an inflated hyperelastic tube:
		theory and experiment}.
	\newblock {\it \bibinfo{journal}{Proc. Roy. Soc. A}\/},  {\it
		\bibinfo{volume}{477}\/}, \bibinfo{pages}{20200837}.
	\bibitem[{Knowles \& Sternberg(1976)}]{ks1976}
	\bibinfo{author}{Knowles, J.~K.}, \& \bibinfo{author}{Sternberg, E.}
	(\bibinfo{year}{1976}).
	\newblock \bibinfo{title}{On the failure of ellipticity of the equations for
		finite elastostatic plane strain}.
	\newblock {\it \bibinfo{journal}{Arch. Ratl Mech. Anal.}\/},  {\it
		\bibinfo{volume}{63}\/}, \bibinfo{pages}{321--336}.
	\bibitem[{Kumar et~al.(2022)Kumar, Audoly \& Lestringant}]{kumar2022asymptotic}
	\bibinfo{author}{Kumar, A.}, \bibinfo{author}{Audoly, B.}, \&
	\bibinfo{author}{Lestringant, C.} (\bibinfo{year}{2022}).
	\newblock \bibinfo{title}{Asymptotic derivation of a higher-order
		one-dimensional model for tape springs}.
	\newblock {\it \bibinfo{journal}{hal-03765944}\/}, .
	\bibitem[{Kyriakides \& Chang(1990)}]{kyriakides1990inflation}
	\bibinfo{author}{Kyriakides, S.}, \& \bibinfo{author}{Chang, Y.-C.}
	(\bibinfo{year}{1990}).
	\newblock \bibinfo{title}{On the inflation of a long elastic tube in the
		presence of axial load}.
	\newblock {\it \bibinfo{journal}{Int. J. Solids Struct.}\/},  {\it
		\bibinfo{volume}{26}\/}, \bibinfo{pages}{975--991}.
	\bibitem[{Kyriakides \& Chang(1991)}]{kyriakides1991initiation}
	\bibinfo{author}{Kyriakides, S.}, \& \bibinfo{author}{Chang, Y.-C.}
	(\bibinfo{year}{1991}).
	\newblock \bibinfo{title}{The initiation and propagation of a localized
		instability in an inflated elastic tube}.
	\newblock {\it \bibinfo{journal}{Int. J. Solids Struct.}\/},  {\it
		\bibinfo{volume}{27}\/}, \bibinfo{pages}{1085--1111}.
	\bibitem[{Lestringant \& Audoly(2018)}]{lestringant2018diffuse}
	\bibinfo{author}{Lestringant, C.}, \& \bibinfo{author}{Audoly, B.}
	(\bibinfo{year}{2018}).
	\newblock \bibinfo{title}{A diffuse interface model for the analysis of
		propagating bulges in cylindrical balloons}.
	\newblock {\it \bibinfo{journal}{Proc. R. Soc. A}\/},  {\it
		\bibinfo{volume}{474}\/}, \bibinfo{pages}{20180333}.
	\bibitem[{Lestringant \&
		Audoly(2020{\natexlab{a}})}]{lestringant2020asymptotically}
	\bibinfo{author}{Lestringant, C.}, \& \bibinfo{author}{Audoly, B.}
	(\bibinfo{year}{2020}{\natexlab{a}}).
	\newblock \bibinfo{title}{Asymptotically exact strain-gradient models for
		nonlinear slender elastic structures: a systematic derivation method}.
	\newblock {\it \bibinfo{journal}{J. Mech. Phys. Solids}\/},  {\it
		\bibinfo{volume}{136}\/}, \bibinfo{pages}{103730}.
	\bibitem[{Lestringant \& Audoly(2020{\natexlab{b}})}]{lestringant2020one}
	\bibinfo{author}{Lestringant, C.}, \& \bibinfo{author}{Audoly, B.}
	(\bibinfo{year}{2020}{\natexlab{b}}).
	\newblock \bibinfo{title}{A one-dimensional model for elasto-capillary
		necking}.
	\newblock {\it \bibinfo{journal}{Proc. R. Soc. A}\/},  {\it
		\bibinfo{volume}{476}\/}, \bibinfo{pages}{20200337}.
	\bibitem[{Lin et~al.(2020)Lin, Li \& Ye}]{lly2020}
	\bibinfo{author}{Lin, Z.~H.}, \bibinfo{author}{Li, L.~A.}, \&
	\bibinfo{author}{Ye, Y.} (\bibinfo{year}{2020}).
	\newblock \bibinfo{title}{Numerical simulation of localized bulging in an
		inflated hyperelastic tube with fixed ends}.
	\newblock {\it \bibinfo{journal}{Int. J. Appl. Mech.}\/},  {\it
		\bibinfo{volume}{12}\/}, \bibinfo{pages}{2050118}.
	\bibitem[{Liu et~al.(2019)Liu, Ye, Althobaiti \& Xie}]{liu2019prevention}
	\bibinfo{author}{Liu, Y.}, \bibinfo{author}{Ye, Y.},
	\bibinfo{author}{Althobaiti, A.}, \& \bibinfo{author}{Xie, Y.-X.}
	(\bibinfo{year}{2019}).
	\newblock \bibinfo{title}{Prevention of localized bulging in an inflated
		bilayer tube}.
	\newblock {\it \bibinfo{journal}{Int. J. Mech. Sci.}\/},  {\it
		\bibinfo{volume}{153}\/}, \bibinfo{pages}{359--368}.
	\bibitem[{Lu et~al.(2015)Lu, An, Li, Yuan \& Wang}]{laly2015}
	\bibinfo{author}{Lu, T.~Q.}, \bibinfo{author}{An, L.}, \bibinfo{author}{Li,
		J.~G.}, \bibinfo{author}{Yuan, C.}, \& \bibinfo{author}{Wang, T.~J.}
	(\bibinfo{year}{2015}).
	\newblock \bibinfo{title}{Electro-mechanical coupling bifurcation and bulging
		propagation in a cylindrical dielectric elastomer tube}.
	\newblock {\it \bibinfo{journal}{J. Mech. Phy. Solids}\/},  {\it
		\bibinfo{volume}{85}\/}, \bibinfo{pages}{160--175}.
	\bibitem[{Lu et~al.(2020)Lu, Ma \& Wang}]{lmw2020}
	\bibinfo{author}{Lu, T.~Q.}, \bibinfo{author}{Ma, C.}, \&
	\bibinfo{author}{Wang, T.~J.} (\bibinfo{year}{2020}).
	\newblock \bibinfo{title}{Mechanics of dielectric elastomer structures: A
		review}.
	\newblock {\it \bibinfo{journal}{Extreme Mech. Lett.}\/},  {\it
		\bibinfo{volume}{38}\/}, \bibinfo{pages}{100752}.
	\bibitem[{Lu \& Suo(2012)}]{ls2012}
	\bibinfo{author}{Lu, T.~Q.}, \& \bibinfo{author}{Suo, Z.~G.}
	(\bibinfo{year}{2012}).
	\newblock \bibinfo{title}{Large conversion of energy in dielectric elastomers
		by electromechanical phase transition}.
	\newblock {\it \bibinfo{journal}{Acta Mech. Sin.}\/},  {\it
		\bibinfo{volume}{28}\/}, \bibinfo{pages}{1106--1114}.
	\bibitem[{Ma et~al.(2015)Ma, Huang, Liu, Li, Qu \& Yang}]{mh2015}
	\bibinfo{author}{Ma, G.~Y.}, \bibinfo{author}{Huang, X.~Q.},
	\bibinfo{author}{Liu, J.~J.}, \bibinfo{author}{Li, T.~F.},
	\bibinfo{author}{Qu, S.~X.}, \& \bibinfo{author}{Yang, W.}
	(\bibinfo{year}{2015}).
	\newblock \bibinfo{title}{Dielectric elastomer peristaltic pump module with
		finite deformation}.
	\newblock {\it \bibinfo{journal}{Smart Mat. Struct.}\/},  {\it
		\bibinfo{volume}{24}\/}, \bibinfo{pages}{075026}.
	\bibitem[{M{\"u}ller et~al.(2008)M{\"u}ller, Lang, Dominietto, Rudin, Schulz,
		Deyhle, Germann, Pfeiffer, David \& Weitkamp}]{muller2008high}
	\bibinfo{author}{M{\"u}ller, B.}, \bibinfo{author}{Lang, S.},
	\bibinfo{author}{Dominietto, M.}, \bibinfo{author}{Rudin, M.},
	\bibinfo{author}{Schulz, G.}, \bibinfo{author}{Deyhle, H.},
	\bibinfo{author}{Germann, M.}, \bibinfo{author}{Pfeiffer, F.},
	\bibinfo{author}{David, C.}, \& \bibinfo{author}{Weitkamp, T.}
	(\bibinfo{year}{2008}).
	\newblock \bibinfo{title}{High-resolution tomographic imaging of microvessels}.
	\newblock In {\it \bibinfo{booktitle}{Developments in X-ray Tomography VI}\/}
	(pp. \bibinfo{pages}{89--98}).
	\newblock \bibinfo{organization}{SPIE} volume \bibinfo{volume}{7078}.
	\bibitem[{Pamplona et~al.(2006)Pamplona, Goncalves \& Lopes}]{pg2006}
	\bibinfo{author}{Pamplona, D.~C.}, \bibinfo{author}{Goncalves, P.~B.}, \&
	\bibinfo{author}{Lopes, S. R.~X.} (\bibinfo{year}{2006}).
	\newblock \bibinfo{title}{Finite deformations of cylindrical membrane under
		internal pressure}.
	\newblock {\it \bibinfo{journal}{Int. J. Mech. Sci.}\/},  {\it
		\bibinfo{volume}{48}\/}, \bibinfo{pages}{683--696}.
	\bibitem[{Pearce \& Fu(2010)}]{pearce2010characterization}
	\bibinfo{author}{Pearce, S.~P.}, \& \bibinfo{author}{Fu, Y.~B.}
	(\bibinfo{year}{2010}).
	\newblock \bibinfo{title}{Characterization and stability of localized
		bulging/necking in inflated membrane tubes}.
	\newblock {\it \bibinfo{journal}{IMA J. Appl. Math.}\/},  {\it
		\bibinfo{volume}{75}\/}, \bibinfo{pages}{581--602}.
	\bibitem[{Pipkin(1968)}]{pipkin1968}
	\bibinfo{author}{Pipkin, A.~C.} (\bibinfo{year}{1968}).
	\newblock \bibinfo{title}{Integration of an equation in membranes theory}.
	\newblock {\it \bibinfo{journal}{Z. Angew. Math. Phys.}\/},  {\it
		\bibinfo{volume}{19}\/}, \bibinfo{pages}{818--819}.
	\bibitem[{Smith(2016)}]{smith2016}
	\bibinfo{author}{Smith, Q.~R.} (\bibinfo{year}{2016}).
	\newblock \bibinfo{title}{Wave-structure interactions for the distensible tube
		wave energy converter}.
	\newblock {\it \bibinfo{journal}{Proc. R. Soc. A}\/},  {\it
		\bibinfo{volume}{472}\/}, \bibinfo{pages}{20160160}.
	\bibitem[{Stano \& Percoco(2021)}]{stano2021additive}
	\bibinfo{author}{Stano, G.}, \& \bibinfo{author}{Percoco, G.}
	(\bibinfo{year}{2021}).
	\newblock \bibinfo{title}{Additive manufacturing aimed to soft robots
		fabrication: A review}.
	\newblock {\it \bibinfo{journal}{Extreme Mech. Lett.}\/},  {\it
		\bibinfo{volume}{42}\/}, \bibinfo{pages}{101079}.
	\bibitem[{Varatharajan \& DasGupta(2017)}]{vd2017}
	\bibinfo{author}{Varatharajan, N.}, \& \bibinfo{author}{DasGupta, A.}
	(\bibinfo{year}{2017}).
	\newblock \bibinfo{title}{Study of bifurcation in a pressurized hyperelastic
		membrane tube enclosed by a soft substrate}.
	\newblock {\it \bibinfo{journal}{Int. J. Non-linear Mech.}\/},  {\it
		\bibinfo{volume}{95}\/}, \bibinfo{pages}{233--241}.
	\bibitem[{Wang et~al.(2017)Wang, Althobaiti \& Fu}]{wang2017localized}
	\bibinfo{author}{Wang, J.}, \bibinfo{author}{Althobaiti, A.}, \&
	\bibinfo{author}{Fu, Y.~B.} (\bibinfo{year}{2017}).
	\newblock \bibinfo{title}{Localized bulging of rotating elastic cylinders and
		tubes}.
	\newblock {\it \bibinfo{journal}{J. Mech. Mater. Struct.}\/},  {\it
		\bibinfo{volume}{12}\/}, \bibinfo{pages}{545--561}.
	\bibitem[{Wang \& Fu(2018)}]{wang2018effect}
	\bibinfo{author}{Wang, J.}, \& \bibinfo{author}{Fu, Y.~B.}
	(\bibinfo{year}{2018}).
	\newblock \bibinfo{title}{Effect of double-fibre reinforcement on localized
		bulging of an inflated cylindrical tube of arbitrary thickness}.
	\newblock {\it \bibinfo{journal}{J. Eng. Math.}\/},  {\it
		\bibinfo{volume}{109}\/}, \bibinfo{pages}{21--30}.
	\bibitem[{Wang \& Fu(2021)}]{wang2021necking}
	\bibinfo{author}{Wang, M.}, \& \bibinfo{author}{Fu, Y.~B.}
	(\bibinfo{year}{2021}).
	\newblock \bibinfo{title}{Necking of a hyperelastic solid cylinder under axial
		stretching: Evaluation of the infinite-length approximation}.
	\newblock {\it \bibinfo{journal}{Int. J. Eng. Sci.}\/},  {\it
		\bibinfo{volume}{159}\/}, \bibinfo{pages}{103432}.
	\bibitem[{Wang et~al.(2022)Wang, Jin \& Fu}]{wjf2022}
	\bibinfo{author}{Wang, M.}, \bibinfo{author}{Jin, L.~S.}, \&
	\bibinfo{author}{Fu, Y.~B.} (\bibinfo{year}{2022}).
	\newblock \bibinfo{title}{Axi-symmetric necking versus treloar-kearsley
		instability in a hyperelastic sheet under equibiaxial stretching}.
	\newblock {\it \bibinfo{journal}{Math. Mech. Solids}\/},  {\it
		\bibinfo{volume}{to appear}\/}.
	\bibitem[{Wang et~al.(2019)Wang, Guo, Zhou, Li \& Fu}]{wang2019experimental}
	\bibinfo{author}{Wang, S.~B.}, \bibinfo{author}{Guo, Z.~M.},
	\bibinfo{author}{Zhou, L.}, \bibinfo{author}{Li, L.~A.}, \&
	\bibinfo{author}{Fu, Y.~B.} (\bibinfo{year}{2019}).
	\newblock \bibinfo{title}{An experimental study of localized bulging in
		inflated cylindrical tubes guided by newly emerged analytical results}.
	\newblock {\it \bibinfo{journal}{J. Mech. Phys. Solids}\/},  {\it
		\bibinfo{volume}{124}\/}, \bibinfo{pages}{536--554}.
	\bibitem[{Wolfram(1991)}]{wo1991}
	\bibinfo{author}{Wolfram, S.} (\bibinfo{year}{1991}).
	\newblock {\it \bibinfo{title}{Mathematica: A System for Doing Mathematics by
			Computer (2nd Edn)}\/}.
	\newblock \bibinfo{publisher}{Addison-Wesley, California}.
	\bibitem[{Ye et~al.(2019)Ye, Liu, Althobaiti \& Xie}]{ye2019localized}
	\bibinfo{author}{Ye, Y.}, \bibinfo{author}{Liu, Y.},
	\bibinfo{author}{Althobaiti, A.}, \& \bibinfo{author}{Xie, Y.-X.}
	(\bibinfo{year}{2019}).
	\newblock \bibinfo{title}{Localized bulging in an inflated bilayer tube of
		arbitrary thickness: Effects of the stiffness ratio and constitutive model}.
	\newblock {\it \bibinfo{journal}{Int. J. Solids Struct.}\/},  {\it
		\bibinfo{volume}{176}\/}, \bibinfo{pages}{173--184}.
	\bibitem[{Ye et~al.(2020)Ye, Liu \& Fu}]{ye2020weakly}
	\bibinfo{author}{Ye, Y.}, \bibinfo{author}{Liu, Y.}, \& \bibinfo{author}{Fu,
		Y.~B.} (\bibinfo{year}{2020}).
	\newblock \bibinfo{title}{Weakly nonlinear analysis of localized bulging of an
		inflated hyperelastic tube of arbitrary wall thickness}.
	\newblock {\it \bibinfo{journal}{J. Mech. Phys. Solids}\/},  {\it
		\bibinfo{volume}{135}\/}, \bibinfo{pages}{103804}.
	\bibitem[{Yin(1977)}]{yin1977}
	\bibinfo{author}{Yin, W.-L.} (\bibinfo{year}{1977}).
	\newblock \bibinfo{title}{Non-uniform inflation of a cylindrical elastic
		membrane and direct determination of the strain energy function}.
	\newblock {\it \bibinfo{journal}{J. Elast.}\/},  {\it \bibinfo{volume}{7}\/},
	\bibinfo{pages}{265--282}.
	\bibitem[{Yu \& Fu(2022)}]{yu2022analytic}
	\bibinfo{author}{Yu, X.}, \& \bibinfo{author}{Fu, Y.~B.}
	(\bibinfo{year}{2022}).
	\newblock \bibinfo{title}{An analytic derivation of the bifurcation conditions
		for localization in hyperelastic tubes and sheets}.
	\newblock {\it \bibinfo{journal}{Z. Angew. Math. Phys.}\/},  {\it
		\bibinfo{volume}{73}\/}, \bibinfo{pages}{1--16}.
	
\end{thebibliography}

\end{document}